%% file: occurrence_in_DR2.tex
\documentclass{aa}
\input{latex_mycommands.txt}

\usepackage{graphicx}
\usepackage{amssymb}
\usepackage{multirow,bigdelim}
\usepackage{txfonts}
\usepackage[export]{adjustbox}
\usepackage{longtable}
\usepackage{lscape}
\usepackage{color}

\usepackage{hyperref}
\hypersetup{colorlinks,citecolor=blue,filecolor=black,linkcolor=red,urlcolor=black}
\begin{document} 

\title{The \planck\ clusters in the \lofar\ sky}
\subtitle{IV: \lotss-DR2: statistics of radio halos and re-acceleration models}

\authorrunning{R. Cassano et al.} 
\titlerunning{The \planck\ clusters in the \lofar\ sky}

\author{R. Cassano\inst{\ref{ira}}, 
V. Cuciti\inst{\ref{hamburg}}, 
G. Brunetti\inst{\ref{ira}}, 
A. Botteon\inst{\ref{unibo},\ref{ira},\ref{leiden}}, 
M. Rossetti\inst{\ref{iasf}}, 
L. Bruno\inst{\ref{ira},\ref{unibo}},
A. Simionescu\inst{\ref{sron},\ref{leiden},\ref{kavli}}, 
F. Gastaldello\inst{\ref{iasf}},
R. J. van Weeren\inst{\ref{leiden}}, 
M. Br\"uggen\inst{\ref{hamburg}},
D. Dallacasa\inst{\ref{unibo},\ref{ira}}
X. Zhang\inst{\ref{leiden},\ref{sron}}, 
H. Akamatsu\inst{\ref{sron}}, 
A. Bonafede\inst{\ref{unibo},\ref{ira}}, 
G. Di Gennaro\inst{\ref{hamburg}}, 
T. W. Shimwell\inst{\ref{astron},\ref{leiden}}, 
F. de Gasperin\inst{\ref{ira},\ref{hamburg}},
H. J. A. R\"ottgering\inst{\ref{leiden}} and
A. Jones\inst{\ref{hamburg}}
}

\institute{
INAF - IRA, via P.~Gobetti 101, I-40129 Bologna, Italy 
\label{ira}\\
\email{rcassano@ira.inaf.it} 
\and
Hamburger Sternwarte, Universit\"{a}t Hamburg, Gojenbergsweg 112, D-21029 Hamburg, Germany \label{hamburg}
\and
Dipartimento di Fisica e Astronomia, Universit\`{a} di Bologna, via P.~Gobetti 93/2, I-40129 Bologna, Italy 
\label{unibo}
\and
Leiden Observatory, Leiden University, PO Box 9513, NL-2300 RA Leiden, The Netherlands \label{leiden}
\and
INAF - IASF Milano, via A.~Corti 12, I-20133 Milano, Italy \label{iasf}
\and
SRON Netherlands Institute for Space Research, Niels Bohrweg 4, NL-2333 CA Leiden, The Netherlands  \label{sron}
\and
Kavli Institute for the Physics and Mathematics of the Universe, The University of Tokyo, Kashiwa, Chiba 277-8583, Japan\label{kavli}
\and
ASTRON, the Netherlands Institute for Radio Astronomy, Postbus 2, NL-7990 AA Dwingeloo, The Netherlands \label{astron}
}

\date{Received XXX; accepted YYY}

\abstract
{Diffuse cluster-scale synchrotron radio emission is discovered in an increasing number of galaxy clusters in the form of radio halos, probing the presence of relativistic electrons and magnetic fields in the intra-cluster medium (ICM). The favoured scenario to explain their origin is that they trace turbulent regions generated during cluster-cluster mergers where particles are re-accelerated. In this framework, radio halos are expected to probe cluster dynamics and are predicted to be more frequent in massive systems where more energy becomes available to the re-acceleration of relativistic electrons.
For these reasons, statistical studies of galaxy cluster samples have the power to derive fundamental information on the radio halo populations and on their connection with cluster dynamics, and hence to constrain theoretical models. Furthermore, low-frequency cluster surveys have the potential to shed light on the existence of radio halos with very steep radio-spectra, which are a key prediction of turbulent models that should be generated in less energetic merger events and thus be more common in the Universe.}
{The main question we will address in this paper is whether we can explain the observed properties of the RH population in this sample within the framework of current models.
}
{We study the occurrence and properties of radio halos from clusters of the second catalog of \planck\ Sunyaev Zel'dovich detected sources that lie within the 5634 deg$^2$ covered by the second Data Release of the \lotssE. We derive their integral number, flux density and redshift distributions. We compare these observations with expectations of theoretical models. We also study the connection between radio halos and cluster mergers by using cluster morphological parameters derived through Chandra and/or XMM-Newton data.
}
{We find that the number of observed radio halos, their radio flux density and redshift distributions are in line with what is expected in the framework of the re-acceleration scenario. In line with model expectations, the fraction of clusters with radio halos increases with the cluster mass, confirming the leading role of the gravitational process of cluster formation in the generation of radio halos. These models predict a large fraction of radio halos with very steep spectrum in the DR2 Planck sample, this will be tested in future studies, yet a comparison of the occurrence of halos in GMRT and LOFAR samples indeed shows a larger occurrence of halos at lower frequencies suggesting the presence of a population of halos with very steep spectrum that is preferentially detected by LOFAR.
Using morphological information we confirm that radio halos are preferentially located in merging systems and that the fraction of newly LOFAR discovered radio halos is larger in less disturbed systems.} 
{}


\maketitle
%

\section{Introduction}

Galaxy clusters are the largest gravitationally bound structures in the Universe, containing mass of $\sim 10^{14}-10^{15}\,M_{\odot}$ in $2-3$ Mpc sized regions. They are made especially by dark matter (70-80\% of the cluster mass), by a $\sim15-20\%$ of a hot and low density gas,
the intracluster medium (ICM),
and by a few percent of cluster galaxies \citep[\eg][for a review]{kravtsov12rev}.
Observations of galaxy clusters in the radio band show the presence of diffuse synchrotron radiation in merging galaxy clusters in the form of radio halos at the cluster centers and radio relics at the cluster outskirts. Mini-haloes (MH) are also observed as less extended extended objects surrounding the central radio galaxy associated with the brightest cluster galaxy in typically relaxed clusters \citep[\eg][for reviews]{feretti12rev, vanweeren19rev}. A recent evidence is the detection of radio bridges connecting pairs of massive clusters \citep[\eg][]{govoni19,botteon20a1758,brunetti20}.
All these sources prove the existence of cosmic-ray electrons and magnetic fields in the ICM and pose fundamental questions about the origin of these components; their impact on the thermal ICM (microphysics); their connection with the cluster dynamics and evolution \citep[\eg][for a review]{brunetti14rev}. 

Radio Halos (RHs) are diffuse, Mpc-sized, synchrotron radio sources with steep radio spectra ($\alpha>1$, with $f(\nu)\propto \nu^{-\alpha}$) that are observed in the central regions of a fraction of galaxy clusters. A statistical connection between RH and the cluster dynamical status has been found, with clusters with RHs often showing signs of merger activity from the analysis of X-ray observations  \citep[\eg][]{buote01, cassano10connection, wen13, cuciti15, giacintucci17}. These observations provided support to the hypothesis that the turbulence generated during cluster mergers re-accelerates pre-existing fossil and/or secondary electrons in the ICM to the energies necessary to produce the observed radio emission \citep[\eg][]{brunetti14rev}.

According to turbulent re-acceleration models \citep[\eg][]{brunetti01coma,petrosian01,fujita03,cassano05,brunetti07turbulence,brunetti11mfp,beresnyak13,donnert13,miniati15run,brunetti16stochastic,pinzke17,nishiwaki22},  
the formation history of RHs depends on the cluster merging rate throughout cosmic epochs and on the
mass of the hosting clusters themselves, which ultimately sets the energy budget that is available for the acceleration of relativistic particles. 
A statistical model based on semi-analytic calculation of the galaxy clusters formation history \citep[][]{press74,lacey93} and on a simple prescription (\ie homogeneous conditions) to estimate the turbulence injected during mergers and the synchrotron spectra that are generated by the turbulent (re)acceleration process, has been developed and tested in the last decade \citep[\eg][]{cassano05,cassano06,cassano10lofar,cassano12}.
A key expectation of this scenario is that the statistical properties of the population of RHs should depend on the frequency of the observations, since the synchrotron spectra of the halos are characterised by a steepening frequency $\nu_s$. Above this frequency the spectrum of RHs gradually steepens and thus the halo could be difficult to observe at $\nu>\nu_s$. Since the value of $\nu_s$ 
is connected to the energetics of the merger events that generate the halos, observations at $\sim 1$ GHz frequency should discover RHs preferentially in massive objects undergoing energetic merging events. RHs in less massive merging-systems should be difficult to observe at frequency $\geq$ 1 GHz \citep[][]{cassano06,cassano10lofar}. 
Statistical studies of clusters, such as the GMRT Radio Halo Survey \citep[][]{venturi07,venturi08,kale13,kale15}, and its extension to mass-selected clusters \citep[from the Planck SZ cluster catalogue;][]{planck14xxix} led to the first statistical evidence that radio halos are predominately found in merging clusters, whereas clusters without diffuse emission are typically relaxed \citep[\eg][]{brunetti07cr,cassano10connection,cassano13}. They have also shown that radio halos become progressively less common in less massive clusters \citep[\eg][]{cuciti21a,cuciti21b}, thus hinting at an observational connection between radio halo formation and the energetics of the hosting systems.

However, the key expectation of the model that
a large fraction of RH has very steep spectrum ($\alpha>1.5$ at $\sim1$ GHz frequency) and glows up preferentially at lower frequencies remains so far poorly explored.
Yet there was an emerging evidence of halos with very steep spectrum detected at lower frequencies \citep[\eg][]{brunetti08,macario13, wilber18a1132, duchesne21arx2,bruno21,rajpurohit21macsj0717halo} supporting the existence of such a population. Also there are emerging evidences of clusters that exhibit some level of dynamical disturbance 
hosting multi-component halos, a central mini-halo like emission surrounded by a diffuse radio emission on larger scale \citep[\eg][]{savini18group,savini19,biava21,Riseley22}. 
In this respect, \lofar\ has recently enabled observations of galaxy clusters at frequencies $<$200 MHz with unprecedented high-sensitivity and resolution. More specifically,  \lofar\ is carrying out wide and deep surveys of the entire Northern sky
at 120$-$168 MHz and 42$-$66 MHz in the context of the \lotssE\ \citep[\lotss;][]{shimwell17} and \lolssE\ \citep[\lolss;][]{degasperin21}, respectively. One of the main goals of these surveys is the discovery of new diffuse Mpc-scale radio sources in galaxy clusters providing samples suitable to test the formation models.
A first step in this direction has been carried out by \cite{vanweeren21} who performed a first statistical investigation of diffuse emission in galaxy clusters selected from the second \planck\ catalog of SZ sources \citep[PSZ2;][]{planck16xxvii} that have been covered by the first \lotss\ Data Release \citep[\lotss-DR1;][]{shimwell19}. A further important step to constrain the spectrum of RH
has been carried out by \cite{digennaro21fast,digennaro21highz}, who observed a small sample of massive clusters at high redshift ($z\geq0.6$) with LOFAR and then followed up the detected halos with the uGMRT. In line with models, about 50\% of these radio halos exhibit a very steep spectral index (\ie\ $\alpha \geq 1.5$ between 150-650 MHz) and are found among the less massive clusters in that sample.

More recently, we have started a large project\footnote{Images, tables, and further information of all targets can be found on the project website \url{https://lofar-surveys.org/planck_dr2.html}.} aiming at the study of diffuse radio emission in the ICM of the galaxy clusters selected from the PSZ2 clusters that have been covered by the second \lotss\ Data Release \citep[\lotss-DR2;][]{shimwell22}, covering 5634 square degrees (27\% of the northern sky).
In \citet[][Paper I]{Botteon22} we present the sample, describe the methods and data used, classify the cluster radio sources, and provide measurements of different quantities. In \citet{bruno22sub} we present the procedures and derive the upper limits to the radio flux density and power of clusters where we do not observe diffuse emission. In \cite{Zhang22} we derive the X-ray properties for clusters in the sample with available \chandra\, and/or \xmm\, archival data. In Cuciti et al. (in prep.) we study the scaling relations of radio halos in the sample, comparing the distribution of radio halos and upper limits in the cluster mass vs radio power diagram. In \cite{jones23} we present the statistical analysis of radio relics in the same sample. In this paper, we focus on radio halos and investigate their flux density and redshift distribution, their occurrence as a function of the cluster mass and redshift, their connection with the cluster dynamics and compare these findings with the expectations from theoretical models.

\indent
Hereafter, we adopt a \lcdm\ cosmology with $\omegal = 0.7$, $\omegam = 0.3$, and $\hzero = 70$ \kmsmpc.

\begin{figure}
\centering
\includegraphics[width=\hsize,trim={0cm 0cm 0cm 0cm},clip]{{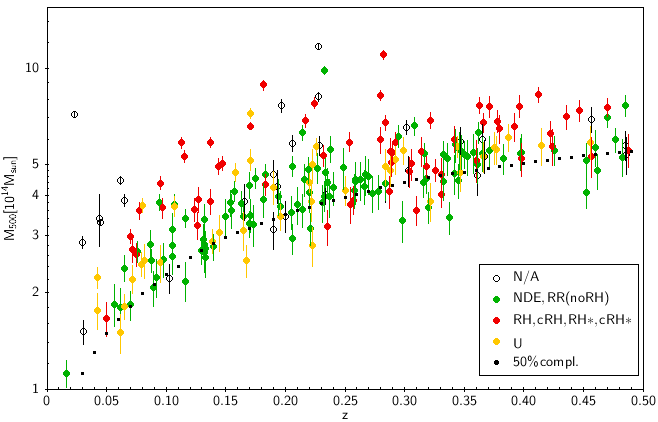}}
\caption{The redshift-mass distribution of PSZ2 clusters within the DR2 area up to $z=0.5$. Different colours show the radio classification (see figure legend).}
\label{fig:dr2_sample}
\end{figure}

\section{Cluster sample}\label{sec:sample}

The PSZ2 catalog \citep{planck16xxvii} contains 1653 SZ-sources detected over the entire sky. Of these 309 lie in the \lotss-DR2 footprint and are the subject of paper I, and 281 out of 309 have a mass and a redshift information. 

In Paper I we classify the diffuse radio emission in those clusters by visually inspecting a set of \lofar\ images at different resolutions (with and without individual point source subtraction) together with the optical/X-ray overlay images. To make this classification most objective and easily reproducible, we make use of a decision tree 
to classify the diffuse emission. Briefly (see Paper I for details), the radio sources are classified as  \textit{Radio halos (RH)}, extended sources that occupy the region where the bulk of the X-ray emission from the ICM is detected;
\textit{Radio relics (RR)}, elongated sources whose position is offset from the bulk of the X-ray emission from the ICM; 
\textit{Candidate radio halos/relics (cRH/cRR)}, a rather clear RH/RR emission is present, but the absence of \chandra\ or \xmm\ X-ray observations do not allow to firmly claimed them;
\textit{Uncertain (U)}, if the emission was either significantly affected by calibration/subtraction artefacts or if it did not fall easily in the categories of radio halos and relics; \textit{No diffuse emission (NDE)}, these objects do not show the presence of diffuse emission that is not associated with AGN; \textit{Not applicable (N/A)}, the emission cannot be adequately classified because of poor data quality.
It is important to note that in these studies we prefer to not divide mini-halos from RHs based on the size of the radio emission, and thus refer generically to radio halos (see Paper I). 

As described in Paper I we used the \halofdcaE\footnote{\url{https://github.com/JortBox/Halo-FDCA}} \citep[\halofdca;][]{boxelaar21} to measure the integrated flux density from the observed radio halos assuming exponential profiles for the fitting. This model has two main parameters: the central surface brightness ($I_0$) and the  \textit{e}-folding radius ($r_e$) which were determined for each radio halo. As suggested by \citet{murgia09}, when calculating the \halofdca\ derived flux densities, we integrated the best-fit models up to a radius of three times the \textit{e}-folding radius. This choice leads to a flux density which is $\sim$80\% of the one that would be obtained by integrating the model up to infinity and is motivated by the fact that halos do not extend indefinitely. Because the integrated flux density measurements are obtained at 144 MHz, the $k$-corrected radio powers at 150 MHz are derived accordingly to the usual formula and assuming $\alpha=1.3$ (see Eq.5 in Paper I); typical values are indeed in the range $\alpha=1-1.5$ and thus our measurements are only marginally affected by the adopted (unknown) radio spectral index \citep[see \eg][]{feretti12rev,vanweeren19rev}. 
For galaxy clusters where no diffuse radio emission is detected we derive upper limits to the radio power of a possible halo by injecting the visibilities of simulated RHs in the LOFAR data. The injection technique and the resulting upper limits to the flux density are reported in \cite{bruno22sub}.

The radio power-mass correlation for cluster RHs and their comparison with upper limits are discussed in Cuciti et al. (in prep.). Here we will make use of the $P_{150 \rm{MHz}}-M_{500}$ best-fit relation in the form:

\begin{equation}
\mathrm{log}\left(\frac{P_{150 \rm{MHz}}}{10^{24.5}\mathrm{W/Hz}}\right)=B~\mathrm{log}\left(\frac{M_{500}}{10^{14.9}\,M_\odot}\right)+A
\label{Eq:PM}
\end{equation}
with $A=1.1\pm0.1$ and $B=3.59\pm0.48$, which has a measured scatter $\sigma_{raw}\sim0.4$ (see Cuciti et al. in prep.).

For the statistical analysis of the RH properties, we focus on sub-samples of clusters spanning a redshift range of $0.07 < z < 0.5$.
The distribution in redshift and mass of the clusters included in our study is shown in Fig.~\ref{fig:dr2_sample}. The lower redshift cut has been made to exclude the very nearby Universe where the cluster statistics is low because of the limited volume.
The high-$z$ cut has been imposed because
the Planck selection is such as to detect only the most massive clusters which are rare at these redshifts, especially in the relatively small DR2 area \citep[see also a discussion on the incompleteness of the redshift information in][]{planck16xxvii}.
To select our sample we also consider the Planck completeness. Specifically, we convert the selection function (available in the Planck archive, for the full survey region and for a signal-to-noise threshold of $4.5$), originally defined in the SZ signal - cluster size plane, into the $M-z$ plane as described in \citet{planck16xxvii}, and find the boundary line above which the detection probability is larger than a given percentage.
As reference we use the 50\% Planck completeness line in $(M_{500}, z)$ (see Fig.\ref{fig:dr2_sample}).
This means that we can have missed 50\% of clusters with mass close to the 50\% completeness line. However as soon as we go up with the mass, the completeness increases so that we have, for instance, an 80\% completeness for clusters with $M_{500}\simeq 5\times 10^{14}\,M_{\odot}$ at $z\sim0.3$.
We do not expect this choice to introduce significant biases in our results since there are no significant differences between the completeness functions for regular and disturbed clusters, as shown by \citet{planck16xxvii}.  

If we take all clusters with $z=0.07-0.5$ and $M_{500}\geq M_{50\%, Planck}(z)$ (excluding the 17 clusters that have been classified as N/A in Paper I) we end up with a sample of 164 clusters
distributed as:
71 NDE (43\%), 55 RH (31\%), 13 RR (8\%) and 25 U (15\%).

Here with RH we refer to both RH and cRH\footnote{Including also cRH* and RH* that are sources for which it was not possible to make a measurement of the radio power 
(see Paper I for details), hence these will not be used to estimate the RH flux density distribution.}, unless indicated differently. We also point out again that we do not distinguish between MHs and RHs (see Paper I) so in the RH category some MH could be present.
The \textit{U} cases will be treated with caution because they might host a RH or different diffuse sources not necessarily associated with the ICM. Considering also RR clusters with U emission at their center, the total number of U cases we have in the sample is 30.

\begin{figure}
\centering
\includegraphics[width=\hsize,trim={0cm 0cm 0cm 0cm},clip]{{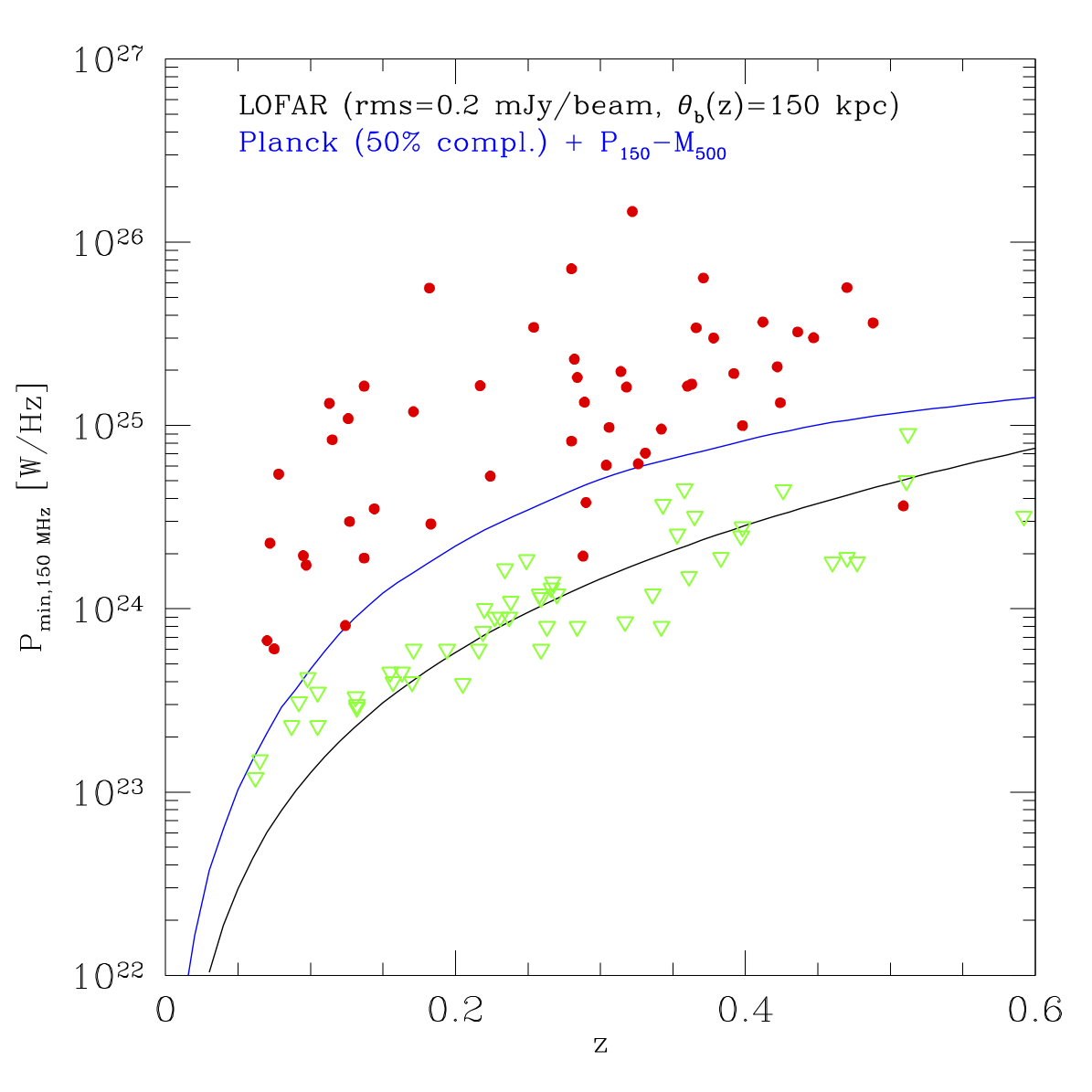}}
\caption{Radio power of halos (red points) and upper limits (green triangles) as a function of redshift. 
The minimum radio power derived from Eq.~\ref{Eq.fmin} is shown as black line (parameters are as in figure panel). The blue line has been obtained by combining the $P_{150 \rm{MHz}}-M_{500}$ best-fit correlation and the 50\% Planck completeness ($M,z$) line in Fig.~\ref{fig:dr2_sample}.}
\label{Fig.Lrmin_z}
\end{figure}

\section{Detecting radio halos in LoTSS DR2}\label{sec:FRH}

In Fig.\ref{Fig.Lrmin_z} we show either the radio power (red points) or upper limits (green triangles) of halos as a function of redshift. This plot allows to explore which is the minimum power of a radio halo that could be detected in LoTSS-DR2. We report also an analytic expression which we used in previous papers \citep[][]{cassano10lofar,cassano12} to estimate the minimum flux density of a RH that can be detected in a given survey by assuming that the halo is detectable
when the integrated flux within $2\times \theta_e$ ($\theta_e$ being the angular size corresponding to the $e-$folding radius $r_e$) gives a signal to noise ratio $\xi$, \ie $f_{min}(<2\,\theta_e)\simeq 0.75 f_{min}(<3\,\theta_e) \simeq \xi\,\sqrt{N_b}\times F_{rms}$, where $N_{b}$ is the number of independent beams within $2\,\theta_e$, it follows:

\begin{equation}
f_{min}(<3\,\theta_e,z)\simeq4.44\times10^{-3}\, \xi\,
\left(\frac{\mathrm{F_{rms}}}{10 \mu\mathrm{Jy}}\right)
\left(\frac{10\,\mathrm{arcsec}}{\theta_b}\right)
\left(\frac{\theta_{e}(z)}{\mathrm{arcsec}}\right)\, \, [\mathrm{mJy}]
\label{Eq.fmin}
\end{equation}

\noindent where $F_{rms}$ is the rms noise in $\mu$Jy
and $\theta_b$ is the beam angular size in arcsec.
The corresponding minimum radio power $P_{min}(z)$ is reported in Fig.~\ref{Fig.Lrmin_z} as black line assuming $F_{rms}=200\,\mu$Jy/beam, $\theta_b(z)$ depending on redshift with a fixed linear size of 150 kpc (see the data reduction strategy described in Paper I) and $\theta_e$ corresponding to $r_e =170$ kpc (which is about the median values of $r_e$ in our sample). With this choice of parameters, Eq.~\ref{Eq.fmin} with $\xi=5$ roughly describes the behaviour of the upper limits
as a function of redshift. The blue line in Fig.~\ref{Fig.Lrmin_z} has been obtained by applying the $P_{150 \rm{MHz}}-M_{500}$ best-fit relation to the 50\% Planck completeness line reported in Fig.~\ref{fig:dr2_sample}. It indicates the minimum power of radio halos in PSZ2 clusters under the assumption that they follow the radio power-mass correlation. The fact that this line is always above the one traced by the upper limits indicates that LOFAR would be able to detect radio halos in clusters with mass above 
the 50\% completeness line. As a consequence, to compare model expectations with observations we will use the blue line to determine which is at each redshift the minimum power of a detectable RH in PSZ2 clusters that lie above the 50\% completeness line (see Sect.\ref{sec:NH}, for details).

\section{Merger-driven turbulent re-acceleration scenario}\label{sec:FRH}

Our LOFAR sample spans a range of cluster masses and redshifts that have never been probed by observations (Fig.~\ref{fig:dr2_sample}). Furthermore it is the first large statistical sample observed at low frequencies, for these reasons it is ideal to test models.
The main goal of the paper is to check if the observed properties of the RH population in this sample
is consistent with model expectations. To explore this point we consider a scenario in which radio halos form in galaxy clusters during cluster-cluster mergers due to the turbulent re-acceleration of relativistic electrons. 

\subsection{Model basic}

A detailed description of the model that we will use can be found in \citet[][]{cassano05} and \citet[][]{cassano06}, while applications to RH predictions for future surveys (with LOFAR, Apertif on WSRT, and ASKAP) can be found in \citet[][]{cassano10lofar,cassano12}. 
In this Section we provide a summary of the theoretical framework and of the most important implications for RH statistical properties and connection with the host clusters.

We model the properties of the RHs and their cosmic evolution by means of a Monte-Carlo approach, which is based on the semi-analytic model of \citet[][\ie, {\it the extended} Press \& Schechter 1974]{lacey93} to describe the hierarchical process of formation of galaxy cluster dark matter halos. The merger history of a synthetic population of galaxy clusters is followed back in time and the generation of turbulence in the ICM is estimated for each merger identified in the {\it merger trees}. In these calculations, turbulence is assumed to be injected in the cluster volume swept by the subclusters, which is bound by the effect of the ram-pressure stripping. The turbulent energy is calculated as a fraction $\eta_t$ ($\sim 0.1-0.3$) of the $P dV$ work done by the subclusters falling into the main cluster.
In these models the turbulent energy, acceleration rate and magnetic field per unit volume are considered constant \citep[i.e. homogeneous models,][]{cassano10lofar}. 

The most important expectation of turbulent re-acceleration scenarios is that the synchrotron spectrum of RHs should become gradually steeper above a frequency, $\nu_s$\footnote{At higher frequencies the synchrotron spectrum of halos steepens. Following \citet[][]{cassano10lofar} we adopt the convention that RHs have spectral index $\alpha=1.9$ between $\nu_s/2.5$ and  $\nu_s$.}, that is determined by the competition between acceleration and energy losses and which is connected to the energetics of the merger events that generate the halos \citep[\eg\,][]{fujita03,cassano05}. In homogeneous models the frequency $\nu_s$ depends on the acceleration efficiency $\chi$, and on the mean magnetic field strength in the radio halo volume $\langle B \rangle$, as $\nu_s\propto \langle B \rangle\,\chi^2 /(\langle B \rangle^2+B_{cmb}^2)^2$ \citep[\eg][]{cassano06,cassano10lofar}, where $B_{cmb}=3.2 (1+z)^2 \mu$G is the equivalent magnetic field of the cosmic microwave background (CMB) radiation. Monte-Carlo simulations of cluster mergers allow us to evaluate $\chi$ from the estimated rate of turbulence-generation and the physical conditions in the ICM. We can then derive the dependence of $\nu_s$ on cluster mass, redshift, and merger parameters in a statistical sample of synthetic clusters \citep[see \eg; ][for details]{cassano05}.

The three main model parameters are $\eta_t$, the typical radius of RH, $R_H$, and the magnetic field in the RH volume $\langle B \rangle$; the dependence of model expectations on the parameter values is explored in a number of papers \citep[\eg][]{cassano06,cassano08revised}. In this paper we limit ourselves to adopt a {\it reference} set of model parameters, namely a volume average magnetic field strength $\langle B \rangle=2\, \mu$G \citep[in line with \eg, ][]{bonafede10}, indipendent on cluster redshift, $\eta_t=0.2$ and a RH size $R_H\simeq 400$ kpc (which is in line with the median size of the halos in our sample). This is the set of parameters that has been routinely used in recent papers \citep[][]{cassano19,botteon21ant,digennaro21highz} and found to reproduce the observed RH statistics at low (LOFAR)
and high (uGMRT) radio frequencies in a sample of high-z clusters  \citep[][]{digennaro21highz}. Furthermore, based on our previous works on the RH statistics at 1.4 GHz \citep[\eg][]{cassano06} and on the comparison between LOFAR and uGMRT RH statistics for high-z clusters \citep[\eg][]{digennaro21highz} we expect that the general results of the present paper are independent of the adopted parameter values. An exploration of the full range of model parameters will be performed upon completion of the LoTSS.

\begin{figure*}
\centering
\includegraphics[width=.33\hsize,trim={0cm 0cm 0cm 0cm},clip,valign=c]{{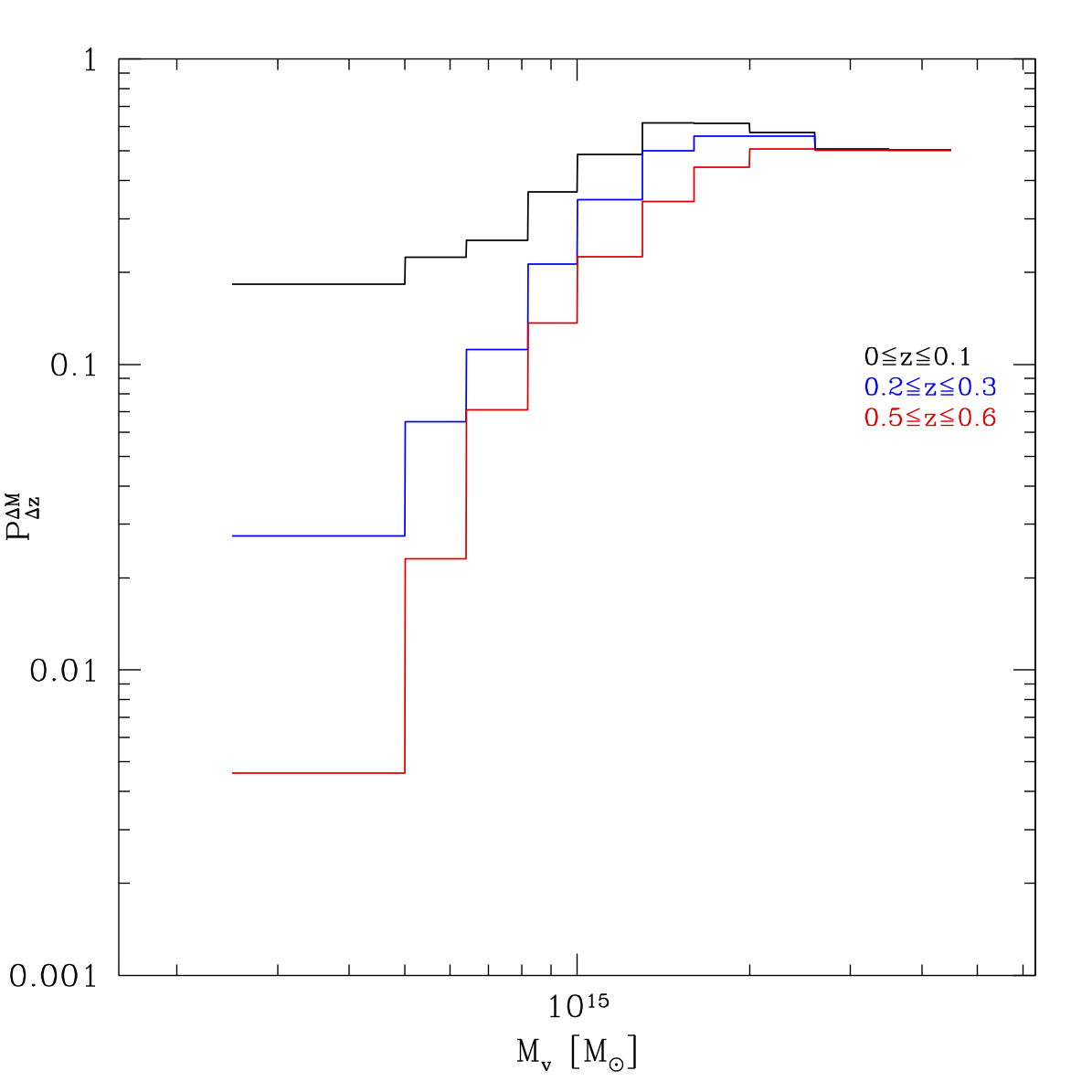}}
\includegraphics[width=.33\hsize,trim={0cm 0cm 0cm 0cm},clip,valign=c]{{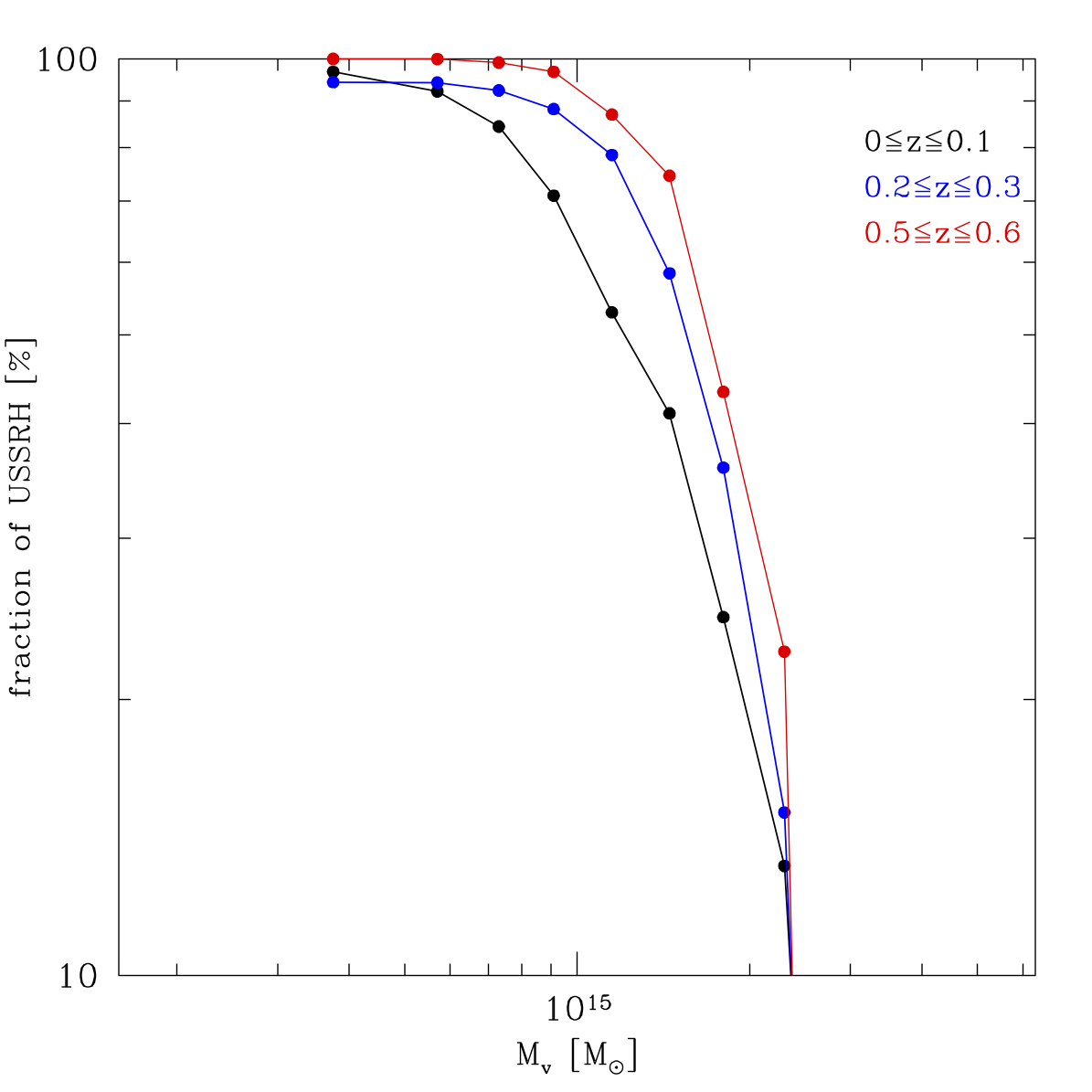}}
\includegraphics[width=.33\hsize,trim={0cm 0cm 0cm 0cm},clip,valign=c]{{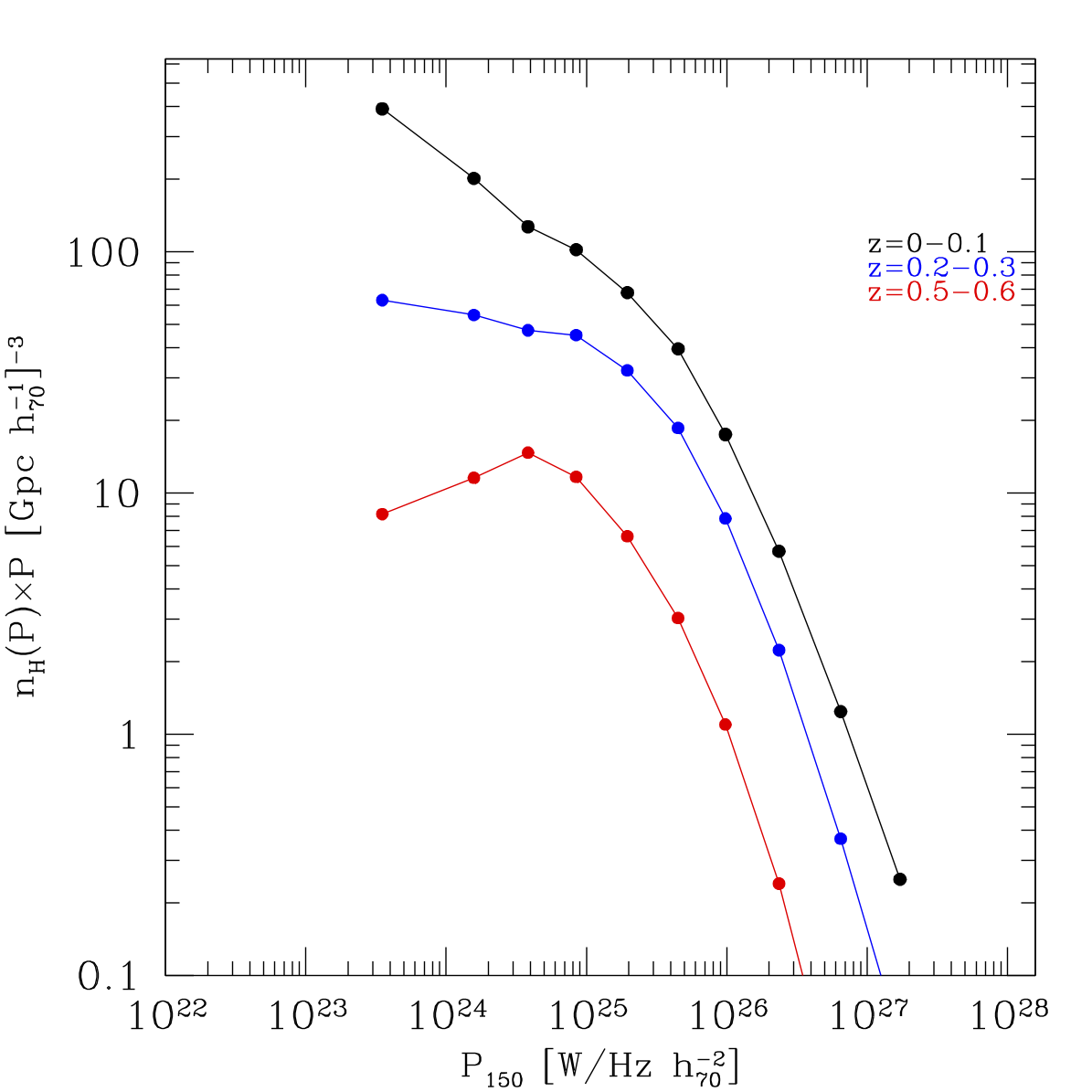}}
\caption{{\it Left panel}: Expected fraction of clusters with radio halos with $\nu_s \geq$ 150 MHz as a function of the cluster mass.{\it Central panel} expected fraction of RH with very steep radio spectra ($\nu_s<600$ MHz) as a function of the cluster mass. {\it Right panel} Radio halo luminosity function at $\nu_o$=150 MHz. In all panels lines refers to z=$0-0.1$ (black line), $0.2-0.3$ (blue line) and $0.5-0.6$ (red line).
}
\label{Fig.fRH}
\end{figure*}

\subsection{Occurrence of radio halos}
As a consequence of the adopted scenario the population of RHs is expected to be made of a complex mixture of sources with different spectra, with massive (and hot) clusters showing the tendency to generate halos with spectra that are flatter than those in less massive systems.

In order to estimate the occurrence of RHs in a survey at a given observing frequency $\nu_o$ we assume that only those halos with $\nu_s \geq \nu_o$ can be observable. Energy arguments imply that RHs with $\nu_s \geq$ 1 GHz are generated in connection with the most energetic merger-events in the Universe. Only these mergers can generate enough turbulence on Mpc-scales and potentially produce the acceleration rate that is necessary to maintain the relativistic electrons emitting at frequencies higher than 1 GHz \citep[][]{cassano05}. In general, in this model the fraction of clusters with radio halos increases with the cluster mass, since more massive clusters are more turbulent \citep[\eg,][]{vazza06,hallman11}, and thus are more likely to host a RH. 
This is in line with the fact that present surveys carried out at $\nu_o \sim$ 1 GHz detect RHs only in the most massive and merging clusters \citep[\eg,][]{cassano13,cuciti15,cuciti21b}. 
For similar energy arguments, RHs with lower values of $\nu_s$, \ie~ or that are with ultra-steep radio spectra (USSRH\footnote{Operatively, in this paper we define USSRH as those halos with $\alpha>1.9$ between 250-600 MHz.}), must be more common since they can be generated in connection with less energetic phenomena, \eg\, major mergers between less massive systems or minor mergers in massive systems, that are more common in the Universe.

\noindent In \cite{cassano12} we show that the fraction of clusters with halos increases at lower values of $\nu_o$, and the size of this increment depends on the considered mass and redshift of the parent clusters, being larger for lower cluster masses and at higher redshifts.

In Fig.~\ref{Fig.fRH} ({\it left panel}), we plot the expected fraction of radio halos with $\nu_s \geq$ 150 MHz (black upper line) as a function of the cluster virial mass and at different redshift (see figure legend and caption).
This is obtained by assuming the {\it reference} set of model parameters (namely $\langle B \rangle=2\, \mu$G, $\eta_t=0.2$). We see that at each redshift the fraction of clusters hosting radio halos with $\nu_s \geq$ 150 MHz increases with the cluster mass. In Fig.~\ref{Fig.fRH} ({\it central panel}), we report the fraction of RHs with $\nu_s \geq$ 150 MHz that in homogeneous models would have very steep radio spectra, specifically those that have $150<\nu_s<600$ MHz. We see that the percentage of very steep spectrum RH is a strong function of the cluster mass, it decreases rapidly for high-mass clusters, and depends also on cluster's redshift. For instance, we found that
in clusters with virial mass $\sim8\times 10^{14}\,M_{\odot}$ at $z\simeq0.05$ (\ie\, $M_{500}\sim 4\times 10^{14}\,M_{\odot}$) $\sim 80-90\%$ of clusters with $\nu_s \geq$ 150 MHz RH has a steep radio spectrum, while at the same redshift this percentage becomes $40\%$ for clusters with $M_v\sim 1.4\times 10^{15}\,M_{\odot}$ (\ie\, $M_{500}\sim 7\times 10^{14}\,M_{\odot}$). For the same masses the percentages of very steep spectrum RH increase considering higher redshifts.

\subsection{The radio halo luminosity function}
\label{Sect.RHLF}

The luminosity functions of radio halos (RHLFs) with $\nu_s\geq \nu_0$ 
(\ie\, the expected number of halos per comoving volume and radio power ``observable'' at
frequency $\nu_0$) 
can be estimated by :

\begin{equation}
{dN_{H}(z)\over{dV\,dP(\nu_0)}}=
{dN_{H}(z)\over{dM\,dV}}\bigg/ {dP(\nu_0)\over dM}\,,
\label{RHLF}
\end{equation}

\noindent
where $dN_{H}(z)/dM\,dV$ is the theoretical mass function of radio 
halos with $\nu_s \geq \nu_0$, that is obtained by combining Monte-Carlo
calculations of the fraction of clusters with RHs and the Press \& Schechter (PS) mass function 
of clusters \citep[\eg\,][]{cassano06}.
We estimate $dP(\nu_0)/dM$ from the radio power-mass correlation,
however with respect to previous papers this is taken directly from the new correlation obtained at
150 MHz for the LOFAR discovered RH (Eq.\ref{Eq:PM} and Cuciti et al. in prep.). The RHLFs at three different redshifts is reported in Fig.~\ref{Fig.fRH} (right panel).

As already discussed in \cite{cassano06} and \cite{cassano10lofar}, the shape of the RHLF flattens at low radio powers because of the expected decrease of the efficiency of particle acceleration in the case of less massive clusters. However, the flattening at low-power is
less relevant than that expected considering higher values of $\nu_0$ \citep[see][]{cassano05} because of the presence of RHs with lower values
of $\nu_s$ contributing to the low power end of the low-frequency luminosity function.   
Finally, we note that the normalisation of the RHLFs decreases with increasing redshift (Fig.~\ref{Fig.fRH}, right panel) due to the evolution with $z$ of both the cluster mass function and the fraction of galaxy clusters with RHs \citep[Fig.~\ref{Fig.fRH}, left panel; see also][]{cassano06}.

\section{Comparison between the observed radio halos in LoTSS DR2 and model predictions}
\label{sec:NH}

In this Section we compare model expectations and the data from LoTSS DR2. As already mentioned we will use a reference values  of model parameters ($\langle B \rangle=2\, \mu$G, $\eta_t=0.2$, $R_H\simeq 400$ kpc) that has been already used in several previous papers (Sect.\ref{sec:FRH}).

The Planck selection function \citep[][see Fig.\ref{fig:dr2_sample}]{planck16xxvii} implies that our measures arise from the combined effect of a simultaneous mass and redshift selection.
In order to proceed with a comparison between observations and models, model predictions must be calculated including the same selection effects we have in the observed sample.

The number of RHs with $flux\geq f_{min}(z)$ in the redshift interval, $\Delta z=z_2-z_1$, can be obtained by integrating the RHLF (Eq.\ref{RHLF}) above a given $f_{min}(z)$\footnote{For a similar approach to radio relics see \cite{bruggen20}.}:

\begin{equation}
N_{H}^{\Delta_z}=\int_{z=z_1}^{z=z_2}dz' ({{dV}\over{dz'}})
\int_{P_{min}(f_{min}^{*},z')}{{dN_H(P(\nu_o),z')}\over{dP(\nu_o)\,dV}}
dP(\nu_o)
\label{Eq.RHNC}
\end{equation}

\noindent
The estimate of $f_{min}(z)$ was the more critical aspect in previous papers where we derived the number of RHs expected to be discovered in future radio surveys. The strength of the present approach is that now we know the capability of LOFAR to detect RHs in PSZ2 clusters (see Sect. 2, for details) because we derived upper limits to the radio flux density of NDE clusters (see Sect. 2 and Bruno et al., submitted). In Sect.~2 we show that the radio power of these upper limits as a function of redshift can be well described by Eq.~\ref{Eq.fmin} and that this power is always lower than the radio power expected for galaxy clusters with $M_{500}\geq M_{50\%, Planck}(z)$ hosting a RH on the $P_{150}-M_{500}$ correlation (\ie\, the blue line in Fig.\ref{Fig.Lrmin_z}). In other words, the Planck selection function ($M_{500}\geq M_{50\%, Planck}(z)$) combined with the $P_{150}-M_{500}$ correlation (Eq.~\ref{Eq:PM}, see also Cuciti et al. in prep) would determine for each redshift which is the minimum radio power of a halo on the correlation we can have in our sample.

\begin{figure}
\centering
\includegraphics[width=\hsize,trim={0cm 0cm 0cm 0cm},clip]{{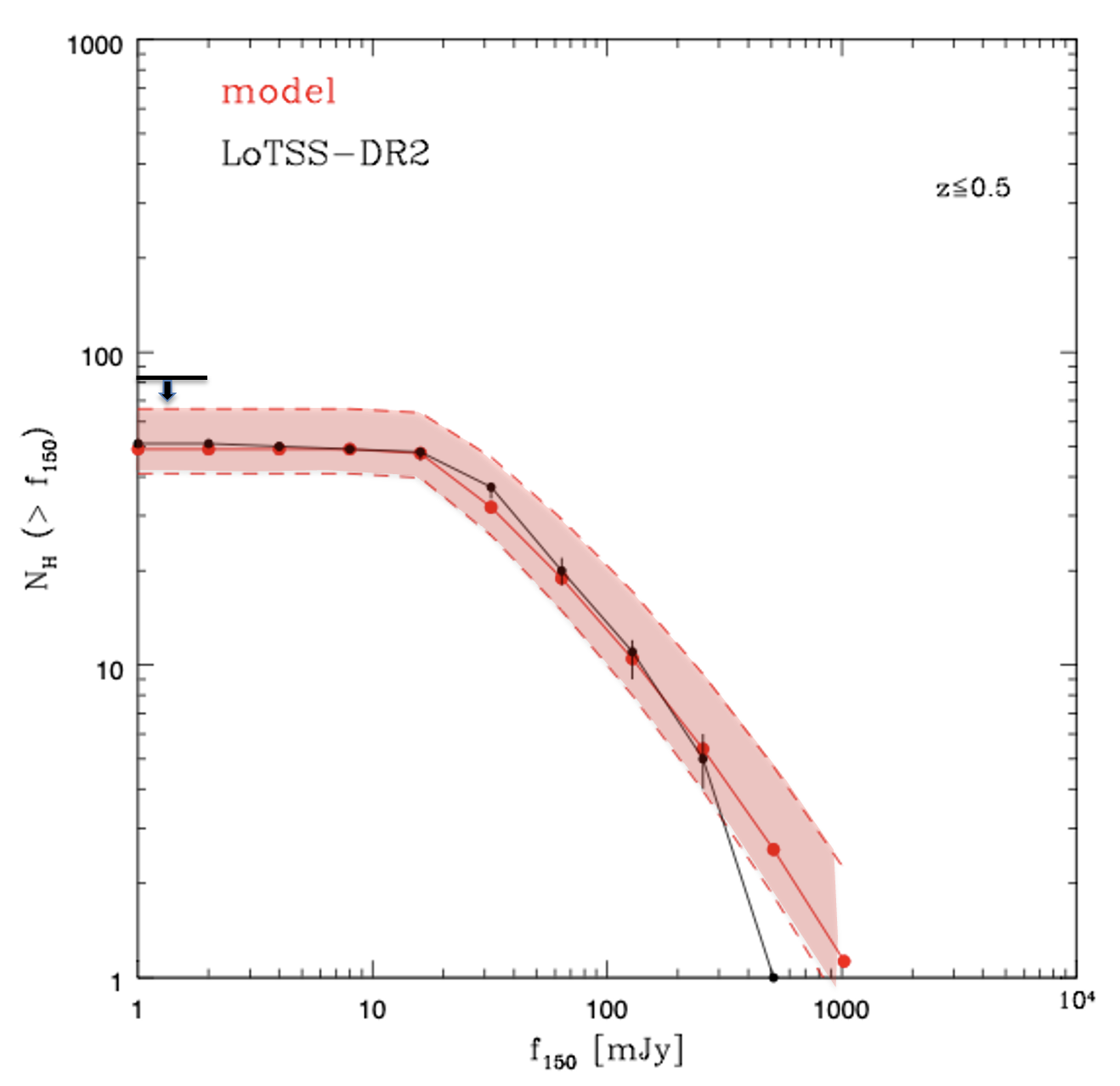}}
\caption{Expected (red line and region) and observed (black dots) number of RHs in PSZ2 clusters with a flux density greater than $f_{150}$ and within $z\leq0.5$ in the \lotss-DR2 area. The black arrow shows the upper boundary to the total number of RHs (due to classification uncertainty), while bars on the black dots are obtained by a Monte Carlo procedure and represent the errors due to uncertainties on the flux densities of RHs.}
\label{Fig.NHf}
\end{figure}

\begin{figure*}
\centering
\includegraphics[width=.42\hsize,trim={0cm 0cm 0cm 0cm},clip,valign=c]{{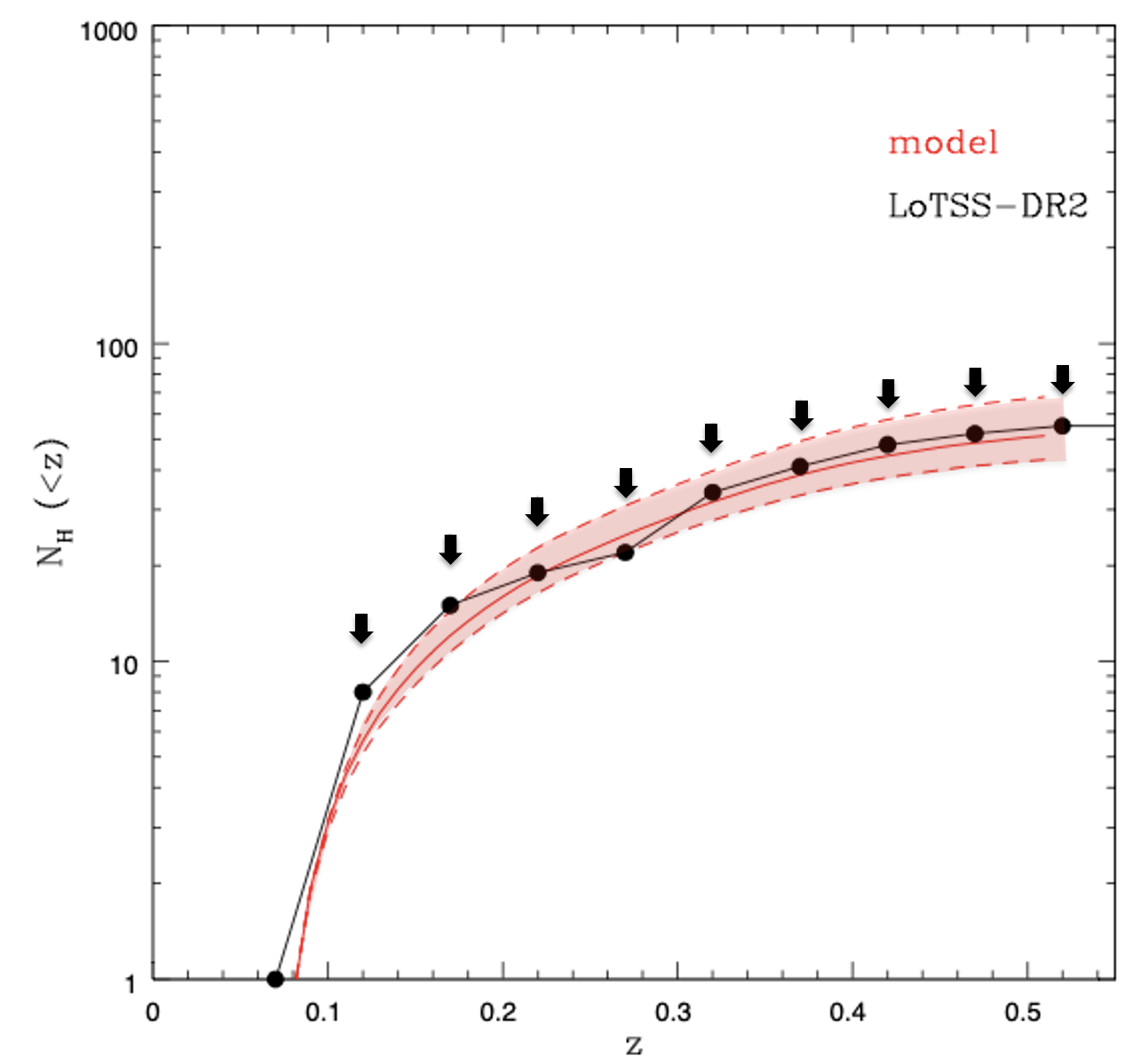}}
\includegraphics[width=.42\hsize,trim={0cm 0cm 0cm 0cm},clip,valign=c]{{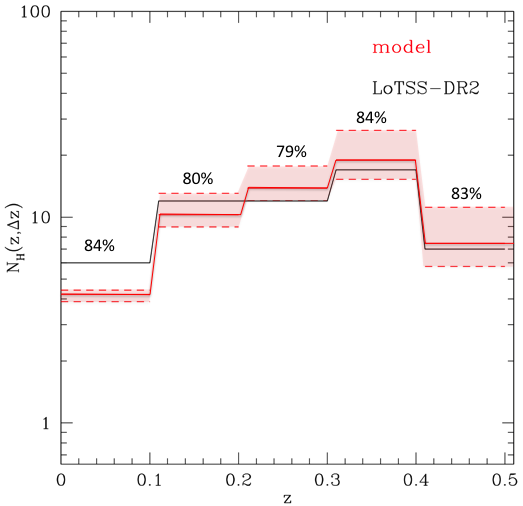}}
\caption{{\it Left panel} integral number of RHs within a given redshift $N_H(<z)$. The black arrows show the upper boundaries to the total number of RH within that redshift bin that are obtained by adding RH and U cases. {\it Right panel} number of RHs in redshift bins $N_H(z,\Delta z)$; the percentage of RHs with ultra-steep spectra (\ie $\nu_s<610 MHz$) is reported in each redshift bin. In both panels, red lines and regions are the expected values, black line and dots the observed ones.}
\label{Fig.NHz}
\end{figure*}

To derive the expected number of LOFAR detectable RHs in PSZ2 clusters in the \lotss-DR2 footprint we use Eq.~\ref{Eq.RHNC} with $f_{min}(z)$ derived as explained above (blue line in Fig.~\ref{Fig.Lrmin_z}). 
In order to compare the predicted number of radio halos with the observed one we first need to normalise the total number of clusters we have in the model (from the Press \& Schechter mass function) so that this matches the number of PSZ2 clusters in our observations.
To do that we divided the predicted number of RHs by the ratio of the number of clusters in the model to the observed number of PSZ2 clusters above the 50\% completeness line.

In Fig.\ref{Fig.NHf} we show the cumulative ($z\leq0.5$) number of RHs expected in PSZ2 clusters in the 
\lotss-DR2 area as a function of the radio halo flux density (red line and region) vs the observed number of RHs (black dots and line). The upper and lower dashed red lines in Fig.\ref{Fig.NHf} define the uncertainties on the expected RH counts and have been obtained by making a Monte Carlo randomisation on the scatter of the $P_{150}-M_{500}$ correlation (see Sect.\ref{sec:sample}).

There are two main sources of uncertainty in the computation of the observed number of RH as function of the radio flux density: {\it i)} flux errors (see Botteon et al 2022), {\it ii)} uncertainty in the RH classification (see Sect.\ref{sec:FRH_mass}).
The latter has been included in Fig.\ref{Fig.NHf} by reporting the upper boundary to the number of RHs that is obtained by simply adding the numbers of RHs and U cases. 
Uncertainties driven by flux errors are modest. They are reported in Fig.\ref{Fig.NHf} using a Monte Carlo procedure which calculates the probability of belonging to the different flux bins, given the statistical and calibration errors of the flux of each RH.

We note a very good agreement between the observed and expected flux density distribution of radio halos\footnote{The deviation of the observed points at the highest flux density could be due to the insufficient statistics, just one observed RH.}.  

The number density of RHs in Fig.~\ref{Fig.NHf} is due to the contribution of RHs observed at different redshifts. Both IC losses and the merging rate for a given mass depend on redsihft, thus it is important to compare model predictions with the observed distribution of RH with $z$.
By using the same normalisation procedure described above, we derive the redshift distributions. In Fig.~\ref{Fig.NHz} we show the cumulative number of RHs within a given redshift $N_H(<z)$ (left panel) and the number of RHs per redshift bins $N_H(z,\Delta z)$ (right panel). In Fig.\ref{Fig.NHz}, left panel, we also report the the upper boundaries to the total number of RHs within a given redshift bin that is obtained by simply adding the numbers of RHs and U.
In both plots we found a very good agreement between the observed and expected distribution of RHs.

\subsection{Radio Halos in different redshift and mass bins: mass dependence}\label{sec:FRH_mass}

\begin{figure*}
\centering
\includegraphics[width=.5\hsize,trim={0cm 0cm 0cm 0cm},clip,valign=c]{{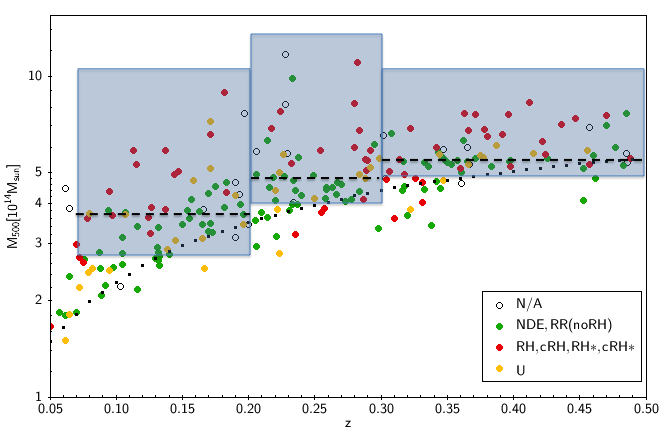}}
\includegraphics[width=.42\hsize,trim={0cm 0cm 0cm 0cm},clip,valign=c]{{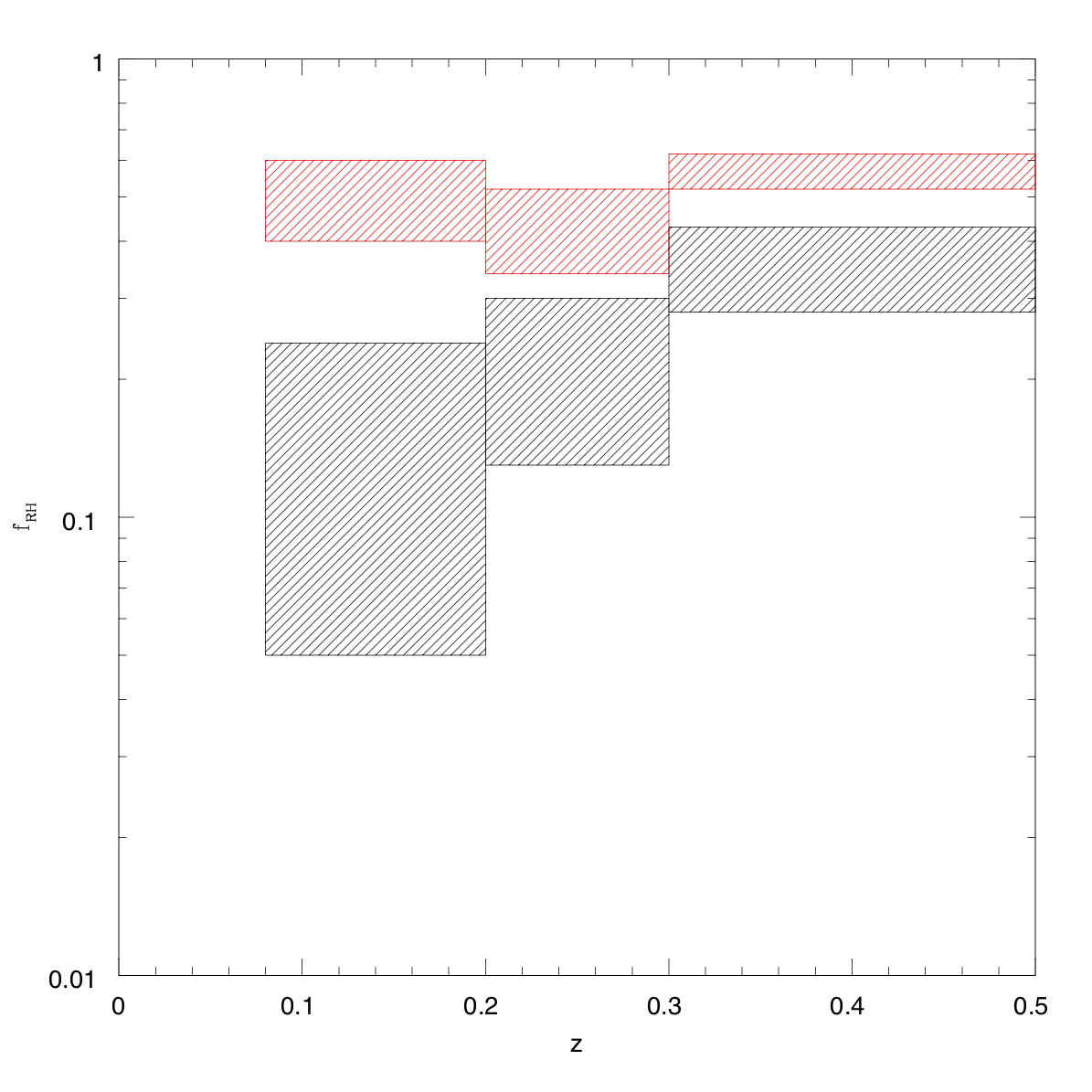}}
\caption{{\it Left panel}: $M_{500}$ vs $z$ distribution of the cluster sample. The rectangular regions show the three redshift bins, and within each of them the dashed line divides the sample into high-mass (above) and low-mass (below) subsamples.
{\it Right panel}: fraction of clusters with radio halos, $f_{RH}$, derived in the high-mass sample (red shadowed region) and in the low-mass sample (black shadowed region). For each bin the reported minimum and maximum values of $f_{RH}$ are $f_{RH}(MC)$ with U=no RH case and $f_{RH}(MC)$ with U= RH case, respectively (see Tab.~\ref{Tab:fraction}).}
\label{fig:FRH_binMz}
\end{figure*}

As already outlined in Sect.~4 the fraction of clusters with RHs is expected to increase with cluster mass in the framework of the particle acceleration scenario, independently of the selected redshift range (see Fig.\ref{Fig.fRH}). In this Section we investigate a dependence of the RH occurrence on the cluster mass in the PSZ2 clusters in the \lotss-DR2 area. This is a difficult task because we need to preserve a high statistics and at the same time consider that the Planck selection (see the black-dotted line in Fig.\ref{fig:dr2_sample}) leads to larger masses at larger redshifts; this implies that our data provide a constraint deriving from a combination of mass and redshift. 
We divide the cluster sample into three redshift bins. In each redshift bin the sample was divided in two
subsamples with similar number of clusters: a high-mass sample and a low-mass sample (see Fig.~\ref{fig:FRH_binMz}, left panel\footnote{Note that, due to the horizontal cuts in mass selection, some clusters just below the 50\% line of completeness can be included in the sample to derive the halo fraction in the various ($M,z$) bins.}).
In Tab.~\ref{Tab:fraction} we report the observed fraction of clusters with RHs, $f_{RH}(obs)$, in the three redshift ranges for both the high-mass and low-mass subsamples.
We show both the cases in which U clusters in the sample are considered as clusters without RH (U= no RH, Tab.~\ref{Tab:fraction}, upper panel) or as RH clusters (U=RH, Tab.~\ref{Tab:fraction}, lower panel).

To take into account the effect on the RH fraction due to the uncertainty associated with the statistical error on the masses we run a Monte Carlo routine. We randomly extract the mass of each cluster from a Gaussian distribution with median value $\mu=M_{500}$ and standard deviation $\sigma$ = $\sigma_{M_{500}}$ , where $\sigma_{M_{500}}$ is the value of the uncertainty on the mass as reported in the Planck catalogue \citep[][]{planck16xxvii}. Then we split the clusters into two sub-samples according to their mass and calculated the fraction of radio halos in each sub-sample. The derived fraction (that is the mean of the distribution) and its uncertainty (that is the standard deviation) are reported in Tab.~\ref{Tab:fraction} as $f_{RH}(MC)$.
In Fig.~\ref{fig:FRH_binMz}, right panel, we show the fraction of clusters with RHs, $f_{RH}$, derived in the high-mass (red shadowed region) and low-mass (black shadowed region) samples in the three redshift bins. For each mass and redshift bins the minimum and maximum reported values are $f_{RH}(MC)$ with U=no RH case and $f_{RH}(MC)$ with U=RH case, respectively (see Tab.~\ref{Tab:fraction}).  
In all three considered redshift ranges it is evident that $f_{RH}$ increases going from the low to the high-mass sample.

\begin{figure*}
\centering
\includegraphics[width=.42\hsize,trim={0cm 0cm 0cm 0cm},clip,valign=c]{{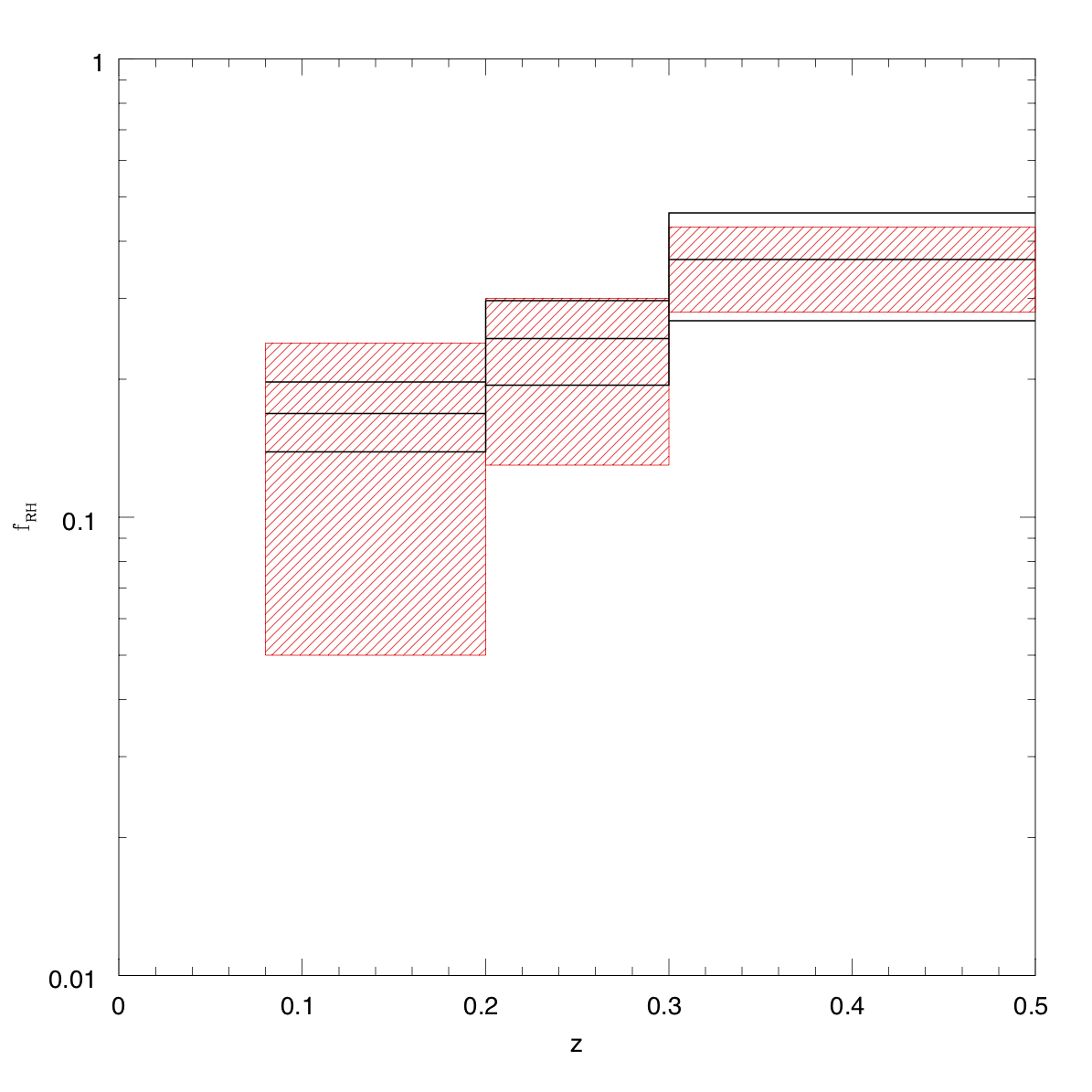}}
\includegraphics[width=.42\hsize,trim={0cm 0cm 0cm 0cm},clip,valign=c]{{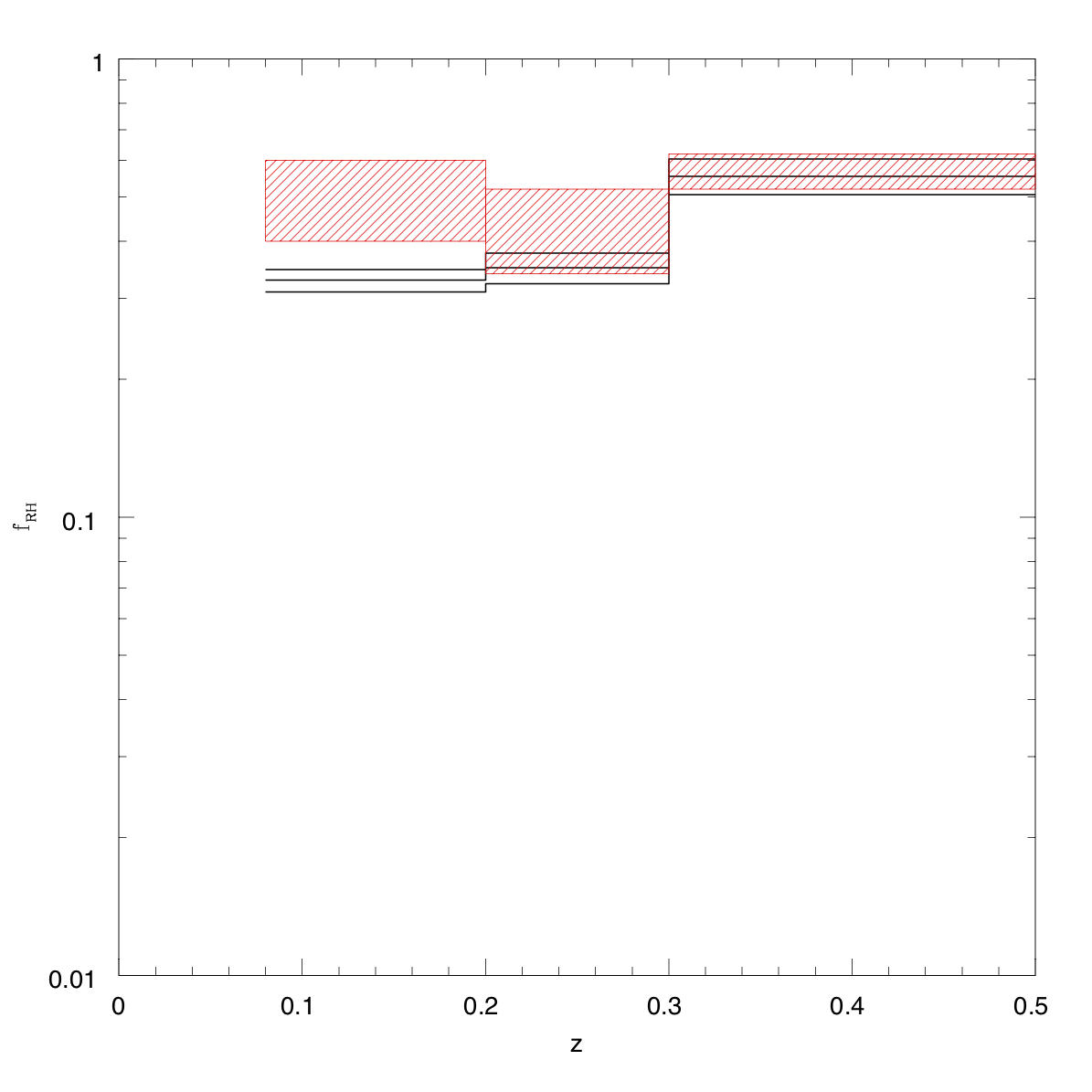}}
\caption{Comparison between expected fractions of clusters with RH (black lines) and observed $f_{RH}$ (red regions) in the three (redshift,mass) bins outlined in Fig.~\ref{fig:FRH_binMz}. Low mass bin values are reported in the left panel, while high-mass bin values in the right panel. For each bin the reported observed minimum and maximum value of $f_{RH}$ are $f_{RH}(MC)$ with U=no RH case and $f_{RH}(MC)$ with U= RH case, respectively (see Tab.~\ref{Tab:fraction}).}
\label{fig:FRH_binMz_theo}
\end{figure*}

We use the same model (as described in Sect. 4) and predict the fraction of clusters with RH in the same mass and redshift bins 
of the observations and report in Fig.~\ref{fig:FRH_binMz_theo} the comparison between the expected and observed values (the same reported also in Fig.~\ref{fig:FRH_binMz}) in the low (left panel) and in the high (right panel) mass bins. We see a fairly solid agreement between expected and derived RH fractions. The model roughly under-reproduces the RH occurrence at low redshift in the high-mass subsample. However this could be explained by the lack of a representative number of massive clusters at low redshift (due to the limited volume of the Universe).

\begin{table*}
\begin{footnotesize}
\begin{center}
\caption{Observed fraction of clusters with radio halos}
\begin{tabular}{|l|cc|cc|cc|}
\hline
&&&&&&\\
	&	\multicolumn{2}{c|}{$z=0.07-0.2$}	&	\multicolumn{2}{c|}{$z=0.2-0.3$}	&	\multicolumn{2}{c|}{$z=0.3-0.5$}\\[2ex]
\hline
&&&&&&\\
	&		$f_{RH}$(obs)	&	$f_{RH}$(MC)	&	$f_{RH}$(obs)	&	$f_{RH}$(MC)	&	$f_{RH}$(obs)	&	$f_{RH}$(MC)\\[2ex]
\hline
&&&&&&\\
U=no RH &  &&&&&\\
&&&&&&\\
high-mass &	$0.37$&	$0.40\pm 0.04$&	$0.36$&	$0.34\pm0.03$&  $0.60$	&	$0.52\pm0.05$\\[3ex]
low-mass 	&	$0.07$&    $0.05\pm 0.03$&	$0.11$&	$0.13\pm0.02$& $0.25$	&	$0.28\pm0.05$\\
&&&&&&\\
\hline
&&&&&&\\
U=RH &  &&&&&\\
&&&&&&\\
high-mass &	$0.59$&	$0.60\pm 0.05$&	$0.56$&	$0.52\pm0.05$&  $0.71$	&	$0.62\pm0.05$\\[3ex]
low-mass 	&	$0.24$	&	$0.24\pm 0.04$&	$0.26$&	$0.29\pm0.04$& $0.35$	&	$0.43\pm0.05$\\
&&&&&&\\
\hline
\end{tabular}	
\label{Tab:fraction}	
\end{center}
\tablefoot {From left to right in each redshift bin the diving mass between the low-mass and high-mass sub-samples are: $M_{500}=3.71\times10^{14}M_\odot$, $4.81\times10^{14}M_\odot$ and $5.68\times10^{14}M_\odot$ (in the U=no RH cases); $M_{500}=3.71\times10^{14}M_\odot$, $4.81\times10^{14}M_\odot$, $5.54\times10^{14}M_\odot$ (in the U=RH cases). Note that the main uncertainty for $f_{RH}(obs)$ is driven by RH classification (upper vs lower panel).}
\end{footnotesize}
\end{table*}

\section{The quest of very steep spectrum halos}
\label{sec:USSRH}

Overall we have shown that a simple version of the re-acceleration models, that is based on homogeneous conditions in the ICM and Monte Carlo simulations of merger-turbulent connection, and that uses reference parameters already adopted in the past, provides an excellent description of the LOFAR observations (Sect. \ref{sec:NH}).

Model predictions are anchored to the assumption that RH can be observed at $\nu_o$ only if $\nu_s \geq \nu_o$.
In particular, we find that the model predicts that most of
halos (70-80\%) detected in PSZ2 clusters in LoTSS-DR2 have steep spectrum ($\nu_s \sim 150-600$ MHz); this is shown in Fig.~\ref{Fig.NHz} ({\it right panel}).
Unfortunately these predictions cannot be tested with current data as many RHs discovered by LOFAR do not have follow up observations at other frequencies.

At the same time in the framework of this simplified model NDE clusters are those systems with $\nu_s < \nu_o$. As a consequence RHs should become visible in a fraction of these systems in sensitive observations at even lower frequencies, for instance in future LOFAR surveys such as \lolss\,  \citep[][]{degasperin21} and LoDeSS (the LOFAR Decameter Sky Survey, van Weeren et al. in prep).

What we can do with current data is to compare the fraction of clusters with RHs observed in our LOFAR sample with that observed with GMRT at 610 MHz in PSZ2 clusters with $\mfive\geq6\times10^{14}\,M_{\odot}$ and $z\leq0.35$ \citep[][]{cuciti21b}. In this mass and redshift range 
we measure a fraction of clusters with RHs $\sim 70 \%$ in our LOFAR sample, compared to $\sim 45 \%$ that is measured in the GMRT sample (see Tab.\ref{Tab:frh_LG}, for details).

\begin{table} 
\begin{footnotesize}
\begin{center}
\caption{Fraction of clusters with radio halos in high-mass clusters}
\begin{tabular}{lcc}
\hline
\hline		
Instrument	&	observed $f_{RH} $	&  $f_{RH}$ from model \\
\hline
LOFAR  & 67-73\%	& $67.0\pm17.8$\%	\\
GMRT & 41-48\%	& $44.6\pm10.0$\%	\\
\hline	
\hline
\end{tabular}	
\label{Tab:frh_LG}	
\end{center}
\tablefoot {Ranges in the observed $f_{RH}$ values are obtained considering the U cases once as RH clusters and once as clusters without RH.}
\end{footnotesize}
\end{table}

\noindent We derive the expected fraction of clusters with RHs in the cluster population with $\mfive\geq 6\times10^{14}\,M_{\odot}$ in the redshift range $z\simeq 0.08-0.35$, by assuming our statistical model (see Sect.\ref{sec:FRH})
and requiring $\nu_s>610$ MHz and $\nu_s>150$ MHz, to be compared with the GMRT and LOFAR fractions, respectively.
We report these fractions in Tab.~\ref{Tab:frh_LG}, together with those derived from the observations.

This can be considered an indirect prove of the presence of USSRH in the PSZ2-DR2 sample. Indeed, we found a good agreement between the observed fractions of clusters with RHs and those derived from the model. The model predicts that the difference between the low and high-frequency fractions is 
caused by the intervening population of very-steep spectrum halos ($\sim 33 \%$ of halos at these cluster masses and redshifts) that become visible preferentially at the lower frequencies.

\begin{figure*}
\centering
\includegraphics[width=.47\hsize,trim={0cm 0cm 0cm 0cm},clip,valign=c]{{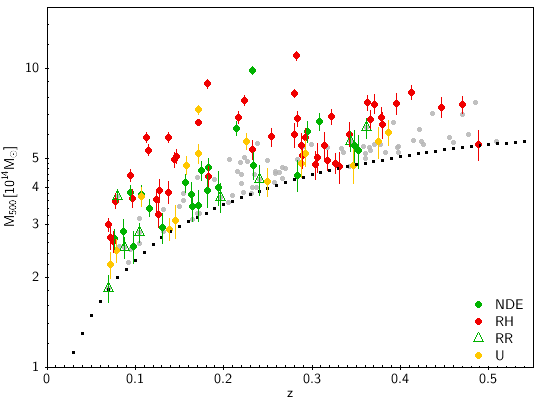}}
\includegraphics[width=.47\hsize,trim={0cm 0cm 0cm 0cm},clip,valign=c]{{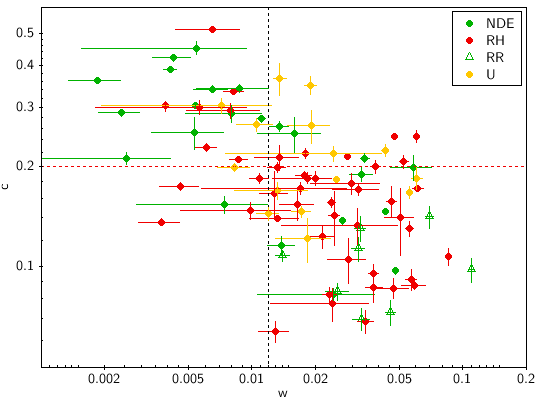}}
\caption{{\it Left panel}: Mass-z distribution of the clusters with $z=0.07-0.5$ above the 50\% Planck completeness line: coloured dots show the radio classification of clusters (see figure legend in the panel)
which have information about their dynamical status; grey dots are the other clusters of the sample without X-ray observations. {\it Right panel}: $c-w$ morphological diagram for the clusters with available X-ray Chandra and/or XMM-Newton data (represented by coloured dots in the {\it Right Panel}). Vertical and horizontal dashed lines are adopted from Cassano et al. (2010) and are: $c=0.2$, $w=0.012$. 
}
\label{fig:morpho_sample}
\end{figure*}

\section{Radio halos and connection with cluster dynamics}\label{sec:morpho}

\begin{table} 
\begin{footnotesize}
\begin{center}
\caption{Relative fraction of NDE and RH clusters in the morphological sample and in the full sample.}
\begin{tabular}{lcc}
\hline
\hline		
sample	& NDE ($f_{NDE}$) 	&  RH ($f_{RH}$)  \\
\hline
full sample  & 71(43\%)	& 55 (31\%)	\\
morpho sample & 23 (24\%)	& 47 (50\%)	\\
\hline
\hline
\end{tabular}	
\label{Tab:cw_morpho}	
\end{center}
\end{footnotesize}
\end{table}

Quantitative measurements of the morphology of the X-ray emission have proved to be an effective way to characterise the dynamical state of large samples of galaxy clusters \citep[e.g.][and references therein]{buote01,santos08, cassano10connection,rasia13rev,rossetti17,lovisari17}. The combination of two morphological parameters as the concentration parameter, $c$ sensitive to the core robustness, and the centroid shift, $w$, sensitive to the presence of substructures, is an optimal choice to characterise the   
merger status of galaxy clusters in relation to their diffuse cluster-scale emission \citep{cassano10connection,cuciti15,cuciti21a}.
In \cite{Zhang22} we derived the morphological parameters over a physical scale of $R_{\rm ap} = 500$ kpc centred on the X-ray emission peak, measurements are reported in 
Paper~I. Here we briefly review the definition of the parameters and then show the results in relation to the analysis of the RH vs NDE clusters.

The concentration parameter has been introduced by \citet{santos08} as the ratio of the photon flux within two circular apertures to effectively identify cool cores even at high redshift. Here we adopt the choice of apertures made by \citet{cassano10connection}

\begin{equation}
 c=\frac{F(r <100\, \rm{kpc})}{F(r <R_{\rm ap})}\:,
\end{equation} 

\noindent
where $F(r <100\, \rm{kpc})$ and $F(r <R_{\rm ap})$ are the exposure-corrected counts within the apertures of 100 kpc and 500 kpc, respectively.

The centroid shift \citep{mohr93, poole06} is defined as the variance of the separation between the X-ray peak and the centroid of the emission obtained within a number of apertures of increasing radius out to $R_{\rm ap}$

\begin{equation}
 w=\left[\frac{1}{N-1}\sum_i(\Delta_i-\overline{\Delta})^2 \right]^{\frac{1}{2}}\frac{1}{R_{\rm ap}}\:,
\end{equation}

\noindent
where $\Delta_i$ is the distance between the X-ray peak and the centroid of the $i$-th aperture. It traces the variation in the position of the centroid introduced by the presence of substructures in the X-ray emission. 
The detailed X-ray data reduction and analysis processes are described in
Paper~I 
and \cite{Zhang22}.

\begin{figure*}
\centering
\includegraphics[width=.42\hsize,trim={0cm 0cm 0cm 0cm},clip,valign=c]{{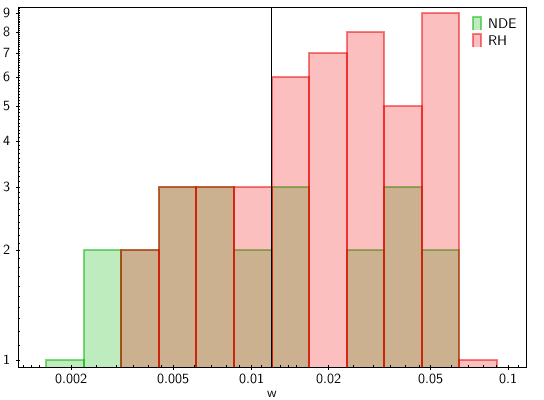}}
\includegraphics[width=.42\hsize,trim={0cm 0cm 0cm 0cm},clip,valign=c]{{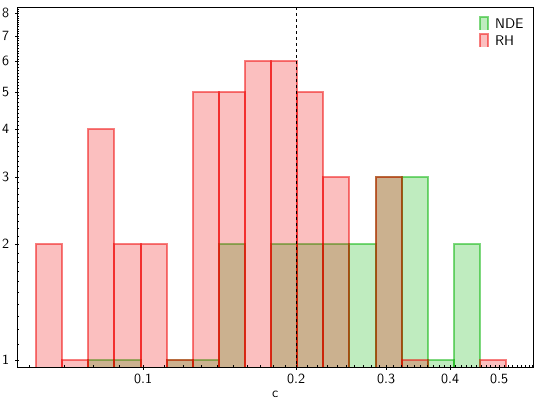}}
\caption{Histogram distributions of $w$ (left panel) and $c$ (right panel) values for clusters in Fig.~\ref{fig:morpho_sample} (coloured dots). Vertical lines are: $c=0.2$, $w=0.012$.}
\label{fig.cw_histo}
\end{figure*}

\begin{figure}
\centering
\includegraphics[width=\hsize,trim={0cm 0cm 0cm 0cm},clip,valign=c]{{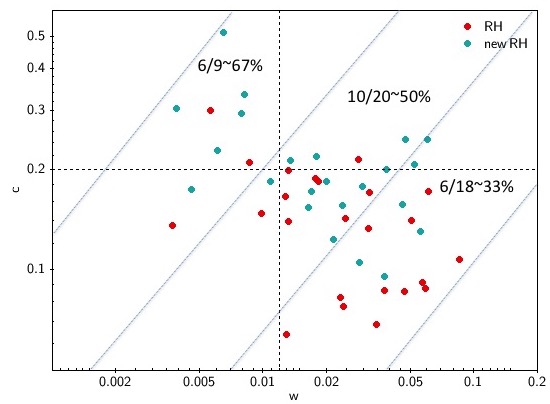}}
\caption{$c-w$ morphological diagram for the clusters with radio halos, new RH, \ie RH detected for the first time by LOFAR are shown with a different colour (cyan dots).
Fractions of newly detected RH are reported in different $c-w$ region. Vertical and horizontal dashed lines are adapted from \citet[][]{cassano10connection} and are: $c=0.2$, $w=0.012$. 
}
\label{Fig.cw_newRH}
\end{figure}

In Fig.~\ref{fig:morpho_sample}, {\it left panel} we report the mass-z distribution of the clusters above the 50\% Planck completeness line with information about their dynamical status (coloured dots) together with the distribution of clusters in the LoTSS-DR2/PSZ2 sample which do not have X-ray observations (grey points)\footnote{We do not report in Fig.\ref{fig:morpho_sample} those clusters which have not been classified in radio, N/A clusters.}. Of the 164 clusters within $z\sim0.07-0.5$ and above the 50\% line, 93 have observations in the \xmm\ and/or \chandra\ archives; here we refer to these clusters as to the ''morphological'' sample. These data have been used to derive the morphological parameters $c$ and $w$ (Paper I, \cite{Zhang22}). When both \xmm\ and \chandra\ measurements are available, we used a value obtained from the combination of the two (see Paper I).
Limitations on the measure of $c$ and $w$ parameters are discussed in \cite{Zhang22}, we conclude that the values of $c$ and $w$ are fairly unbiased since our clusters are below $z=0.5$ and their number of X-ray counts is greater than $1000$.
In Fig.~\ref{fig:morpho_sample}, {\it right panel} we report the $c-w$ distribution of these 93 clusters together with the "reference" dividing lines $c=0.2$ and $w=0.012$ \citep[see][for details]{cassano10connection}. 
We see that the fraction of clusters with RHs has a tendency to increase going from the top left corner (high $c$, low $w$) to the bottom right corner (high $w$, low $c$), in other words going from more relaxed to more disturbed systems. In contrast, the fraction of clusters with NDE tends to increase going towards the more relaxed systems. We also note that, as expected, relic clusters are all in the region of the most disturbed systems
\citep[see][for details]{jones23}.
Unfortunately, the information about the dynamics is incomplete, especially for NDE clusters. Morphological parameters are indeed available for 32\% of NDE; 85\% of RH; 60\% of U; 62\% of RR.
This implies that the relative fractions of RH and NDE clusters in the morphological sample are quite different (50\% and 24\% respectively) from those in the full sample (31\% and 43\%, respectively; see Tab.~\ref{Tab:cw_morpho}). This prevents us from drawing firm conclusions about the relative fraction of RH and NDE vs cluster dynamics. However, since the dynamical information about RHs is quite complete, we can confidently say that RHs are found preferentially in merging systems (see also histogram in Fig.~\ref{fig.cw_histo}), in fact: $\sim77$\% of them live in clusters with $c<0.2$ and $\sim80$\% of them live in clusters with $w>0.012$ ($\sim72\%$ of RH have both $c<0.2$ and $w>0.012$).
In addition, we see that the fraction of newly detected RHs by LOFAR (see cyan points in Fig.~\ref{Fig.cw_newRH}) increases going from the bottom right side (merging) to the upper left side (more relaxed). This indicates that LOFAR is able to unveil radio halos in less disturbed systems. Although we do not have spectral information about these newly detected RHs, based on our framework we speculate that the majority of them are likely characterised by very steep spectra (Sect. \ref{sec:USSRH}; Fig.~\ref{Fig.NHz}). A possible contamination can be due to the presence of MHs in relaxed clusters. However to our knowledge only Abell 1068 (the cluster with the higher value of "c" in Fig.~\ref{fig:morpho_sample}, right panel) is a MH (Biava et al. in prep), additional investigation is required for the other clusters.

\section{Summary}\label{sec:conclusions}

The most commonly accepted scenario to explain the origin of RHs in galaxy clusters assumes that they are the result of particle acceleration due to turbulence produced in the ICM during cluster-cluster mergers. This scenario predicts, besides a connection with cluster mergers, a heterogeneous population of radio halos with synchrotron radio properties, spectral shapes and luminosities that are correlated with the energetics of the merging events. 
For this reason, the study of the statistical properties of the RH populations has the power to constrain their origin and evolution and thus to test theoretical models.

In this paper, we use the Planck clusters in \lotss-DR2 \citep[][]{Botteon22} to carry out the first statistical study of RHs at low radio frequencies.
We select a sub-sample of 164 PSZ2 clusters above the 50\% $(M,z)$ Planck completeness line and spanning a wide redshift $0.07 < z < 0.5$ range. This sample contains $\sim 50$ RH. It allows, for the first time, to make a statistical study of RH in an unprecedented range of cluster masses, including clusters down to $M_{500}\sim 2.5-3\times 10^{14}\,M_{\odot}$, breaking down the wall of $M_{500}\sim 6\times 10^{14}\,M_{\odot}$ that limited previous statistical studies \citep[\eg][]{cassano13,cuciti21a}.
The statistics is sufficient to start investigating the dependence of the RH properties on cluster mass and redshift. However, due to the $(M,z)$ dependence of the Planck selection function, our measurements entail always a mixed information of mass and redshift. 

We compare the occurrence of radio halos in this sample with that derived from a sample of PSZ2 clusters with $M\geq6\times10^{14}\,M_{\odot}$ and $z\leq0.35$ observed with GMRT at 610 MHz \citep{cuciti21b}. In the same mass and redshift range, we find an increase of the fraction of clusters with RHs that goes from $\sim 45 \%$ in the GMRT sample to $\sim 70 \%$ in the LOFAR sample (Sect.~\ref{sec:USSRH}). The observed increase is in line with predictions of the re-acceleration scenario which implies that more RHs should be visible at lower frequency because of their very steep spectra. 
This sample 
offers the unique opportunity to test theoretical models in an uncharted range of mass and redshift and, for the first time, at low radio frequency. 
For this reason,
in our work we have compared quantitatively model expectations with our LOFAR observations. We use semi-analytic models developed in the framework of the merger-driven turbulent re-acceleration scenario \citep[][]{cassano05,cassano06,cassano10lofar} to derive the expected properties of the RH population in the PSZ2 clusters. 
In this paper we limit ourselves to adopt the set of values of model parameters that has been already assumed in a series of previous papers (see Sect.\ref{sec:FRH}).
By using the observed mass and redshift limit of the observed sample and by normalising the number of clusters in the theoretical model to match the observed number of clusters, we show that we can reproduce the integral number of RHs ($\sim 40-70$ expected RH), their flux density and redshift distributions (Sect.~\ref{sec:NH}). Using the same modelling we predict that about 100-200 RH could be detected in PSZ2 clusters (above the 50\% completeness line and $z\sim 0.07-0.5$) by the full LoTSS.

A clear expectation of this model is that the fraction of clusters with RH should increase with the cluster mass. Although this has been already tested at higher frequencies for massive clusters \citep[see \eg][]{cuciti21b}, here we are in the position to investigate for the first time the occurrence of RHs with cluster mass as observed at low radio frequencies and in an unprecedented range of cluster masses. We divided our sample into three redshift bins. For each redshift bin we measure the fraction of clusters with RHs, $f_{RH}$, in the high and low mass sub-samples (Tab.~\ref{Tab:fraction}). We find a clear increase of the RH occurrence with the cluster mass at a fixed redshift. This is in line with expectations derived using the re-acceleration scenario.

Although the statistical model we used is rather simplified, we have shown that it provides an excellent description of the observed properties of the RH population in the PSZ2 clusters in the LoTSS DR2. It reproduces very well their observed integral number, redshift and flux density distributions and it also explains the increase of the RH fraction with the cluster mass and at low radio frequencies. We note that the model, 
with the same set of values of parameters that has been adopted in the current paper, has been shown to explain the RH fraction measured in a sample of massive high-z clusters ($z\simeq 0.6-0.9$) observed with LOFAR and followed-up at higher frequencies with the uGMRT \citep[][]{digennaro21highz}. About 50 \% of the LOFAR detected RHs were found to be characterised by very steep radio spectra, in line with model expectations. 
Future follow-up observations at higher frequency of all RHs of the \lotss-DR2 PSZ2 sample would definitively test this expectation, making the golden test of the scenario. 
Exploiting the full range of model parameters may also provide interesting indications. In the near future, as soon as the full LoTSS is completed, extensive simulations will allow us to identify a meaningful range of values of model parameters that best match the data.

We use the current sample to investigate for the first time the radio halo-cluster merger connection at low radio frequencies. Observations at low radio frequency are expected to find RH also in less disturbed systems. 
Although we have shown that the morphological information is not complete for our sample (especially for NDE clusters) we see that the fraction of clusters with RHs has a tendency to increase 
towards more disturbed systems. In addition, we see that the fraction of newly detected RHs by LOFAR increases going from merging to more relaxed systems, indicating that LOFAR starts to observe RHs in less disturbed systems, possibly unveiling RHs with very steep radio spectra. 
\indent

\begin{acknowledgements}
M.R., R.C., F.G., G.B. acknowledges support from INAF mainstream project "Galaxy Clusters Science with LOFAR". V.C. acknowledges support from the Alexander von Humboldt Foundation. GDG acknowledges support from the Alexander von Humboldt Foundation. And.Bot. and Ann.Bon. acknowledges support from the ERC-StG DRANOEL n. 714245 and from the VIDI research programme with project number 639.042.729, which is financed by the Netherlands Organisation for Scientific Research (NWO). RJvW acknowledges support from the ERC Starting Grant ClusterWeb 804208.
FdG acknowledges support from the Deutsche Forschungsgemeinschaft under Germany's Excellence Strategy - EXC 2121 Quantum Universe - 390833306. A.S. was supported by the Women In Science Excel (WISE) programme of the NWO, and acknowledges the Kavli Institute for the Physics and Mathematics of the Universe for the continued hospitality.
LOFAR \citep{vanhaarlem13} is the LOw Frequency ARray designed and constructed by ASTRON. It has observing, data processing, and data storage facilities in several countries, which are owned by various parties (each with their own funding sources), and are collectively operated by the ILT foundation under a joint scientific policy. The ILT resources have benefitted from the following recent major funding sources: CNRS-INSU, Observatoire de Paris and Universit\'{e} d'Orl\'{e}ans, France; BMBF, MIWF-NRW, MPG, Germany; Science Foundation Ireland (SFI), Department of Business, Enterprise and Innovation (DBEI), Ireland; NWO, The Netherlands; The Science and Technology Facilities Council, UK; Ministry of Science and Higher Education, Poland; Istituto Nazionale di Astrofisica (INAF), Italy. This research made use of the Dutch national e-infrastructure with support of the SURF Cooperative (e-infra 180169) and the LOFAR e-infra group, and of the LOFAR-IT computing infrastructure supported and operated by INAF, and by the Physics Dept.~of Turin University (under the agreement with Consorzio Interuniversitario per la Fisica Spaziale) at the C3S Supercomputing Centre, Italy. The J\"{u}lich LOFAR Long Term Archive and the German LOFAR network are both coordinated and operated by the J\"{u}lich Supercomputing Centre (JSC), and computing resources on the supercomputer JUWELS at JSC were provided by the Gauss Centre for Supercomputing e.V. (grant CHTB00) through the John von Neumann Institute for Computing (NIC). This research made use of the University of Hertfordshire high-performance computing facility and the LOFAR-UK computing facility located at the University of Hertfordshire and supported by STFC [ST/P000096/1]. The scientific results reported in this article are based in part on data obtained from the \chandra\ Data Archive. SRON Netherlands Institute for Space Research is supported financially by the Netherlands Organisation for Scientific Research (NWO). This research made use of APLpy, an open-source plotting package for Python \citep{robitaille12}.
\end{acknowledgements}

%
%

\bibliographystyle{aa}
\bibliography{library.bib}

\end{document}

%% file: latex_mycommands.txt
\def \eg {e.g.}

\def \ie {i.e.}

\def \lcdm {{\hbox{$\Lambda$CDM}}}
\def \omegam {{\hbox{$\Omega_{\rm m}$}}}
\def \omegal {{\hbox{$\Omega_\Lambda$}}}
\def \hzero {{\hbox{$H_0$}}}

\def \mfive {\hbox{$M_{500}$}}

\newcommand{\kmsmpc }{\mbox{km s$^{-1}$ Mpc$^{-1}$}}

\newcommand{\halofdca }{\textsc{Halo-FDCA}}
\newcommand{\halofdcaE }{Halo-Flux Density CAlculator}

\newcommand{\xmm }{{\em XMM-Newton}}
\newcommand{\chandra }{{\em Chandra}}

\newcommand{\planck }{{\em Planck}}

\newcommand{\lofar }{LOFAR}

\newcommand{\lotss }{LoTSS}
\newcommand{\lotssE }{LOFAR Two-meter Sky Survey}
\newcommand{\lolss }{LoLSS}
\newcommand{\lolssE }{LOFAR LBA Sky Survey}